\documentclass[aps,prb,twocolumn,superscriptaddress]{revtex4-2}
\usepackage[utf8]{inputenc}
\usepackage{graphicx, color}
\usepackage{xcolor}
\usepackage{hyperref}
\usepackage{amsmath}
\usepackage{amssymb}
\usepackage{mathtools}
\usepackage{float}
\usepackage{gensymb}
\usepackage{textcomp}
\usepackage{siunitx}
\usepackage{longtable}
\usepackage{natbib}
\usepackage[section]{placeins}
\usepackage{xspace}
\usepackage{url}
\usepackage[toc,page]{appendix}
\usepackage{bm}

\definecolor{link}{rgb}{0.1,0.1,0.9}
\hypersetup{colorlinks=true,linkcolor=link,citecolor=link,urlcolor=link,linktocpage}
\newcommand*\pct{\protect\scalebox{0.9}{\%}\xspace}
\makeatletter
\g@addto@macro\bfseries{\boldmath}
\makeatother

\DeclareRobustCommand{\rchi}{{\mathpalette\irchi\relax}}
\newcommand{\irchi}[2]{\raisebox{\depth}{$#1\chi$}}

\newcommand{\sg}{$R\,\bar{3}\,m$\xspace}
\newcommand{\BXMO}{Ba$_{4}$$\mathcal{M}$Mn$_{3}$O$_{12}$~($\mathcal{M}$~=~Ta, Nb)\xspace}
\newcommand{\BXMOshort}{Ba$_{4}$$\mathcal{M}$Mn$_{3}$O$_{12}$\xspace}
\newcommand{\BNMO}{Ba$_{4}$NbMn$_{3}$O$_{12}$\xspace}
\newcommand{\BTMO}{Ba$_{4}$TaMn$_{3}$O$_{12}$\xspace}
\newcommand{\BNRO}{Ba$_{4}$NbRu$_{3}$O$_{12}$\xspace}

\newcommand{\BXMOSP}{Ba$\mathcal{M}_{0.5}$Mn$_{0.5}$O$_{3}$ ($\mathcal{M}$~=~Ta, Nb)\xspace}

\newcommand{\BRO}{Ba$_{4}$Ru$_{3}$O$_{10}$\xspace}

\newcommand{\BCIO}{Ba$_{5}$CuIr$_{3}$O$_{12}$\xspace}
\newcommand{\MO}{MnO$_{2}$\xspace}
\newcommand{\BNIO}{Ba$_{4}$NbIr$_{3}$O$_{12}$\xspace}
\newcommand{\BNMeO}{Ba$_{4}$Nb$\mathcal{M}$$_{3}$O$_{12}$~($\mathcal{M}$~=~Ru, Ir)\xspace}
\newcommand{\BTMOP}{BaTa$_{0.5}$Mn$_{0.5}$O$_3$\xspace}
\newcommand{\BNMOP}{BaNb$_{0.5}$Mn$_{0.5}$O$_3$\xspace}
\newcommand{\BMMOP}{Ba$\mathcal{M}_{0.5}$Mn$_{0.5}$O$_{3}$~($\mathcal{M}$~=~Ta, Nb)\xspace}

\begin{document}

\preprint{APS/123-QED}
	
\title{Partial molecular orbitals in face-sharing 3$d$ manganese trimer:~\\ Comparative studies on \BTMO and \BNMO}

\author{Anzar Ali}
\thanks{These authors contributed equally to this work.}
\affiliation{Center for Integrated Nanostructure Physics, Institute for Basic Science, Suwon 16419, Republic of Korea}
\affiliation{Sungkyunkwan University, Suwon 16419, Republic of Korea}

\author{Heung-Sik Kim}
\thanks{These authors contributed equally to this work.}
\affiliation{Department of Physics and Institute of Quantum Convergence Technology, Kangwon National University, Chuncheon 24311, Republic of Korea}

\author{Poonam Yadav}
\affiliation{Center for Integrated Nanostructure Physics, Institute for Basic Science, Suwon 16419, Republic of Korea}
\affiliation{Sungkyunkwan University, Suwon 16419, Republic of Korea}

\author{Suheon Lee}
\affiliation{Center for Integrated Nanostructure Physics, Institute for Basic Science, Suwon 16419, Republic of Korea}
\affiliation{Sungkyunkwan University, Suwon 16419, Republic of Korea}

\author{Duhee Yoon}
\affiliation{Center for Integrated Nanostructure Physics, Institute for Basic Science, Suwon 16419, Republic of Korea}
\affiliation{Department of Energy Science, Sungkyunkwan University, Suwon, 16419, Republic of Korea}
	
\author{Sungkyun Choi}
\email{sungkyunchoi@skku.edu}
\affiliation{Center for Integrated Nanostructure Physics, Institute for Basic Science, Suwon 16419, Republic of Korea}
\affiliation{Sungkyunkwan University, Suwon 16419, Republic of Korea}

\begin{abstract}
We present a molecular orbital candidate \BTMO with a face-sharing octahedra trimer, by comparing it with a related compound \BNMO. The synthesis of the polycrystalline powder is optimized by suppressing the secondary impurity phase via x-ray diffraction. Magnetic susceptibility measurements on the optimized samples reveal a weak magnetic hysteresis with magnetic transitions consistent with heat capacity results. The effective magnetic moments from susceptibility indicate a strongly coupled $S=2$ antiferromagnetic trimer at around room temperature, whereas the estimated magnetic entropy from heat capacity suggests the localized $S=3/2$ timer. These results can be explainable by a partial molecular orbital state, in which three $t_{2g}$ electrons are localized in each Mn ion and one $e_{g}$ electron is delocalized over two-end Mn ions of the trimer based on density functional theory calculations. This unconventional 3$d$ orbital state is comprehended as a consequence of competition between the hybrid interatomic orbitals within the Mn trimer and the local moment formation by on-site Coulomb correlations.
\end{abstract}

\maketitle

\section{Introduction}
\label{Intro}
A Mott state~\cite{Imada1998} has been studied extensively for understanding the electronic state of correlated electron systems stabilized by the competition between the electron correlation energy and other energy scales, such as spin-orbit coupling and crystalline electric fields. Quantum magnets containing transition metal (TM) ions can be generally understood in the localized electron scheme by learning the properties of the single magnetic ion and the exchange interactions with its surrounding magnetic ions in a given structural motif.

More recently, new electronic states have attracted significant interest in discovering novel electronic properties. Such an example is a molecular orbital state (MO)~\cite{Streltsov2017}, where electrons are delocalized and shared by several magnetic ions nearby, which cannot be described by a simple single-ion localized electron picture. It has been demonstrated that this unconventional phase can be realized in heavy TM-based compounds, such as the 4$d$ trimer \BRO~\cite{Klein2011, Streltsov2012} and 5$d$ trimer \BCIO~\cite{Ye2018} by the interplay between spin-orbit coupling and extended $d$ orbitals. In these compounds, the TM octahedra form a linear trimer via a face-sharing geometry, which enables an unusually short distance between TM ions, allowing a sizable direct overlap of $d$ orbitals. As a result, they can stabilize the MO state.

However, little is known about the MO state from 3$d$ TM-based compounds. It is generally recognized that characters of 3$d$ orbitals are very different from those of 4$d$ and 5$d$ orbitals. For instance, the spin-orbit coupling  is reduced in 3$d$ orbitals, and electrons are significantly more localized, leading to large electronic Coulomb interactions. In contrast, 3$d$ materials have a higher Hund's coupling energy, which is associated with an intra-atomic exchange in multiorbital systems~\cite{Georges2013}. This distinct energy hierarchy in 3$d$ TM ions compared to those in 4$d$ and 5$d$ counterparts could render the MO state from 3$d$ materials unusual. Therefore, the interplay between spin-orbit and Hund's coupling of 3$d$ compounds could provide the opportunity to discover novel characters of the MO state.

In this regard, a recently reported hexagonal perovskite \BNMO~\cite{Nguyen2019} is intriguing. This compound also contains the linear trimer with a three face-sharing MnO$_{6}$ octahedra, which allows a much closer Mn-Mn distance ($\sim$2.47~\AA) than the one encountered in common corner-sharing geometries~\cite{Poeppelmeier1982}. It is comparable to that of an Mn metal ($\sim$2.48~\AA~\cite{MnMetal}). Thus this structural motif could enable a strong direct overlap between $3d$ orbitals. Nonetheless, \BNMO is an insulator, which suggests an unconventional electronic picture in 3$d$ orbitals. Indeed, magnetic susceptibility measurements uncovered the intriguing magnetic properties of \BNMO~\cite{Nguyen2019}, such as that the effective magnetic moment extracted from high temperatures corresponded to only $S = 2$ from the trimer, which is expected to be a combination of two Mn$^{4+}$ ions and one Mn$^{3+}$ ion. Based on this, a magnetically ordered state within the trimer was suggested even at 300~K~\cite{Nguyen2019} even if the long-range magnetic ordering temperature can be lower than 42.4~K.

To understand the magnetic ground state of \BNMO based on the expected mixed valent state of Mn$^{3+}$ and Mn$^{4+}$, two hypothetical models~\cite{Nguyen2019} in the localized electron scheme are proposed: (i) two parallel moments from Mn$^{4+}$ ions and one antiparallel moment from Mn$^{3+}$ ion (3$d$$^{4}$, $S=1$ for low spin); (ii) two antiparallel moments from Mn$^{4+}$ ions and one moment from Mn$^{3+}$ ion (3$d$$^{4}$, $S=2$ for high spin). Note that the spin of Mn$^{4+}$ ions in both models is the same (3$d$$^{3}$, $S=3/2$).

However, the reported heat capacity~\cite{Nguyen2019} disagrees with both models and reveals a significantly lower magnetic entropy for the $S=2$ system than expected. Thus the consistent characterization of carefully grown samples needs to be done, compared to detailed theoretical calculations using the suggested magnetic structures. In addition, it is crucial to conduct a comparative study to better understand the peculiar electronic properties of \BNMO by synthesizing and examining a related compound if possible.

Herein, we present the synthesis and characterization of a $3d$ trimer \BTMO and an improved sample quality of its related compound \BNMO. The growth optimization is made by controlling an amount of volatile \MO powder, monitored by x-ray diffraction refinements. Sharp phonon peaks observed by Raman scattering corroborate the high quality of our samples. We confirm tiny magnetic hysteresis anomalies and find signatures of magnetic transitions in both compounds in magnetic susceptibility measurements, compatible with heat capacity experiments. The effective magnetic moments from susceptibility indicate a strongly coupled $S=2$ antiferromagnetic trimer at around room temperature, whereas the estimated magnetic entropy from heat capacity suggests the localized $S=3/2$ moments in the timer. {\it Ab initio} calculations find that the localized electrons at Mn $t_{2g}$ orbitals and the delocalized electron at Mn $e_{g}$ orbitals spread over two-end Mn ions of the trimer, which explains both susceptibility and heat capacity results. Hence we propose a partial molecular orbital, which is the coexistent state of localized electrons and molecular orbitals in \BXMO. This peculiar state is interpreted by the result of the competition between the hybrid interatomic orbitals within the Mn trimer and the local moment formation by on-site Coulomb correlations. Further, we discuss the possible relevance of this trimer-based material platform in searching for the insulator-metal transition that might be useful in developing innovative magnetic devices.

The rest of the paper is organized as follows. The experimental and calculational details are shown in Secs.~\ref{Exp} and~\ref{Com}, respectively. The optimization of the polycrystalline powder is explained in Sec.~\ref{sec:growths}. Refined crystal structures by x-ray diffraction are presented in Sec.~\ref{sec:str}. Sections~\ref{sec:Chi} and~\ref{sec:Cp} present magnetic susceptibility and heat capacity results, respectively. Resistivity results are given in Sec.~\ref{sec:Resistivity}. Raman spectra are compared in Sec.~\ref{sec:Raman}, followed by {\it ab initio} calculations in Sec.~\ref{sec:abinitio}. Our results are discussed based on reported references in Sec.~\ref{sec:discussion}, and finally, conclusions are given in Sec.~\ref{sec:conclusions}. The Appendix provides results from nonmagnetic DFT calculations.

\section{Experimental Details}
\label{Exp}
The polycrystalline samples of \BXMO were synthesized using a solid-state reaction. The starting precursors of BaCO$_3$ (Alfa Aesar, 99.95\,\pct), $\mathcal{M}$$_2$O$_5$ ($\mathcal{M}$~=~Ta, Nb)\xspace (Kojundo, 99.9\,\pct), and MnO$_2$ (Kojundo, 99.99\,\pct) were taken in stoichiometric ratio and ground well in an agate mortar using a pestle. Mixed powders were calcined at 900\degree C for 24 h. After the calcination, the powders were repeatedly reground, pelletized, and sintered for 24 hrs at 1100\degree C, 1300\degree C, and 1420\degree C. Slightly modified conditions were also used---the calcination at 1100\degree C, followed by the subsequent sintering at 1200\degree C and 1420\degree C for 24 h at each temperature. Note that our optimized batches were synthesized using the modified conditions. The excess of starting MnO$_2$ powder was also added to compensate for its evaporation, effectively suppressing the impurity phase formation.

The phase purity of polycrystalline powder was confirmed using x-ray diffraction measurements at room temperature using a Rigaku SmartLab X-ray Diffractometer. The high-quality XRD data were collected from a Rigaku D8 advance high-resolution x-ray diffractometer using a Bragg-Brentano geometry with a Cu $K\alpha_{1, 2}$ source, presented in this paper. A {\sc jana}2006 software~\cite{Petricek2014} was used for Rietveld refinements to determine the crystal structures.

Magnetic susceptibility experiments for zero-field-cooled (ZFC) and field-cooled (FC) measurements were performed with a Physical Property Measurement System (PPMS, Dynacool, Quantum Design) equipped with a vibrating sample magnetometry module between 300 and 2~K with a magnetic field of 0.01~T or 0.1~T. Magnetization measurements were done in the magnetic field from $-$7 to 7~T at 2, 30, 50, 70, and 100~K.

Heat capacity measurements were carried out with a PPMS at $H$~=~0 and 9~T from 160 to 2~K. Before each measurement, the addenda was measured using the same temperature range by the thermal relaxation method. Addenda heat capacities were subtracted from total heat capacities to obtain the sample heat capacity as usual.

Electrical resistance was measured on sintered rectangular pellets using a PPMS between 260 and 390~K. The rectangular pellets of 3~$\times$~6 mm$^2$ dimensions were prepared using a rectangular die set. A four-probe method was utilized, where the electrical contacts were made using gold wires and silver epoxy. Measured resistance was converted to resistivity using the length and area of the cross-section.

Raman measurements were performed using the 514.5~nm line of an Ar-ion laser as the excitation source at room temperature. The laser beam was focused onto the sample through a 50$\times$ objective lens (numerical aperture~=~0.8) in a backscattering geometry. The laser power was kept at 0.5~mW to minimize the local beam heating. The Raman signals were recorded with a liquid-nitrogen-cooled charge-coupled-device detector after dispersion by a Horiba iHR550 spectrometer (2400 grooves/mm).

\begin{figure} [t]
\includegraphics[width=\linewidth]{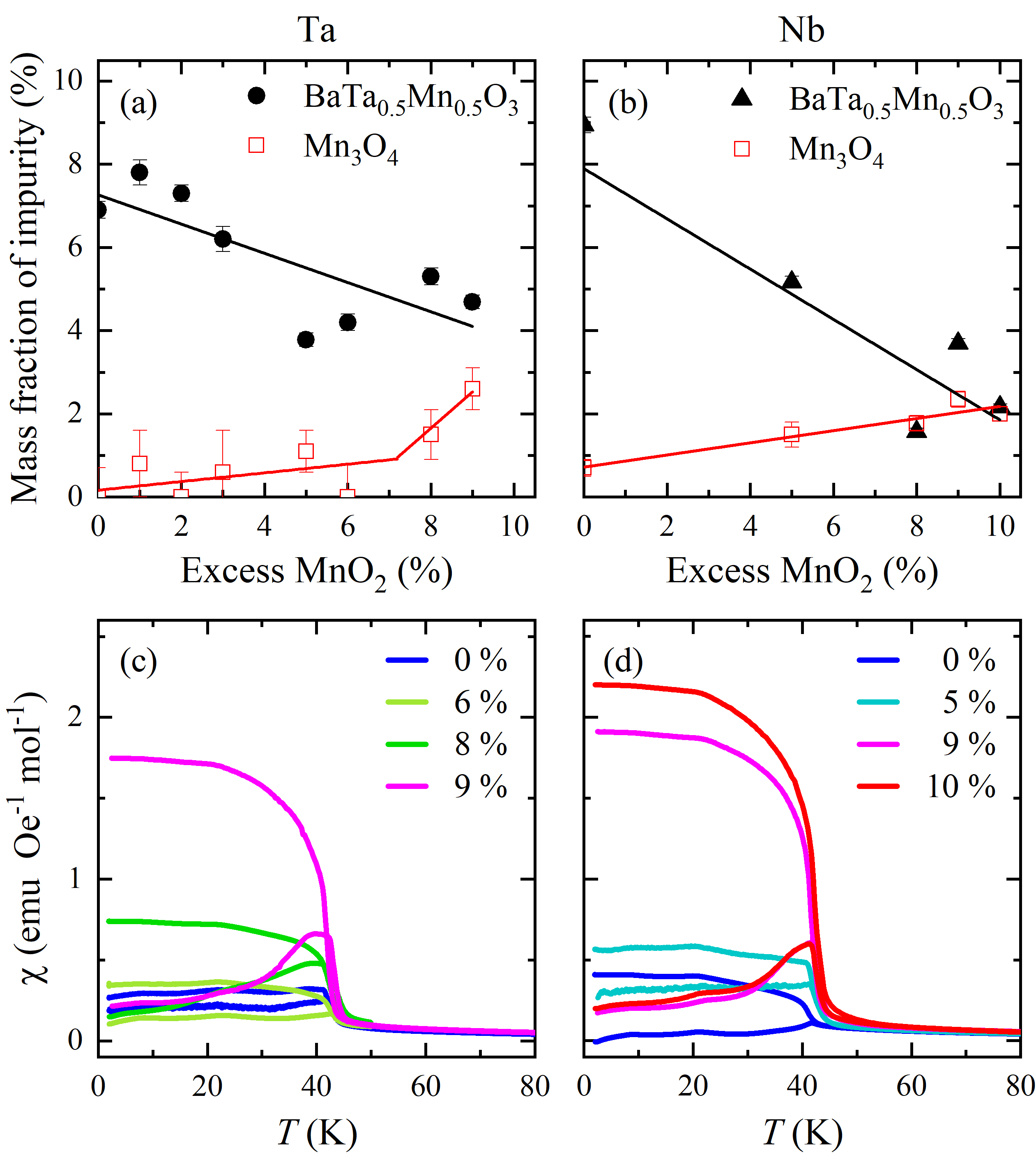}
\caption{Optimization of polycrystalline growths of \BXMO. (a), (b) Evolution of the mass fraction of the unwanted impurity phase, \BTMOP and \BNMOP, and Mn$_{3}$O$_{4}$, upon an excess of \MO starting powder. Error bars from the refinement are also given. (c), (d) Selected magnetic susceptibility for \BTMO and \BNMO at $H$~=~0.01~T, respectively.
}
\label{PF}
\end{figure}

\section{Computational Details}
\label{Com}
The Vienna {\it ab initio} simulation package ({\sc vasp})~\cite{Kresse1993, Kresse1996} was employed for electronic structure calculations, with 400 eV of plane-wave energy cutoff and $5 \times 5 \times 2$ $k$-point sampling for conventional hexagonal unit cells. A revised Perdew-Burke-Ernzherof generalized gradient approximation for solids (PBEsol) was used to approximate the exchange-correlation functional. The strong Coulomb repulsion within the Mn sites was further treated with a simplified rotationally invariant flavor of density functional theory (DFT)+$U_{\rm eff}$~\cite{Dudarev1998}, where the effective on-site Coulomb repulsion $U_{\rm eff} \equiv U - J$ was set to 4 eV for the Mn $d$-orbital. Note that the value of $U_{\rm eff}$ = 4 eV was widely adopted for a range of Mn compounds~\cite{Alex2013,Streltsov2018,HSK2019,Harms2020} and reported to produce reasonable results.

\section{Experiments}

\subsection{Optimization of polycrystalline growths}
\label{sec:growths}
We found the unwelcome impurity phase, \BMMOP, by x-ray diffraction during the synthesis of \BTMO and \BNMO polycrystalline samples. Thus we optimized our growths by controlling the amount of the excess \MO starting powder to compensate for its evaporation during the reaction.

Figure~\ref{PF} presents the mass fraction of the impurity phase extracted by the structural refinements using x-ray diffraction measurements, followed by magnetic susceptibility measurements. As shown in Fig.~\ref{PF}(a), the mass fraction of the impurity phase linearly decreases with an excess of \MO, reaching a minimum value when the 9\% (10\%) of excessive \MO powder is initially added for \BTMO (\BNMO) before the first sintering. We found that the magnetic susceptibility splitting ($\rchi_{\rm FC}$-$\rchi_{\rm ZFC}$) at 2~K is bigger for such batches, as shown in Fig.~\ref{PF}(c) and \ref{PF}(d) for \BTMO and \BNMO, respectively. We studied the structural, magnetic, thermal, electric, and optic properties using the optimized batch of each compound in this paper. We also found a tiny mass fraction of the secondary impurity phase Mn$_{3}$O$_{4}$ ($\sim$2\%) towards a higher excess of MnO$_2$, as shown in Figs.~\ref{PF}(a) and \ref{PF}(b).

\begin{figure} [t]
\includegraphics[width=\linewidth]{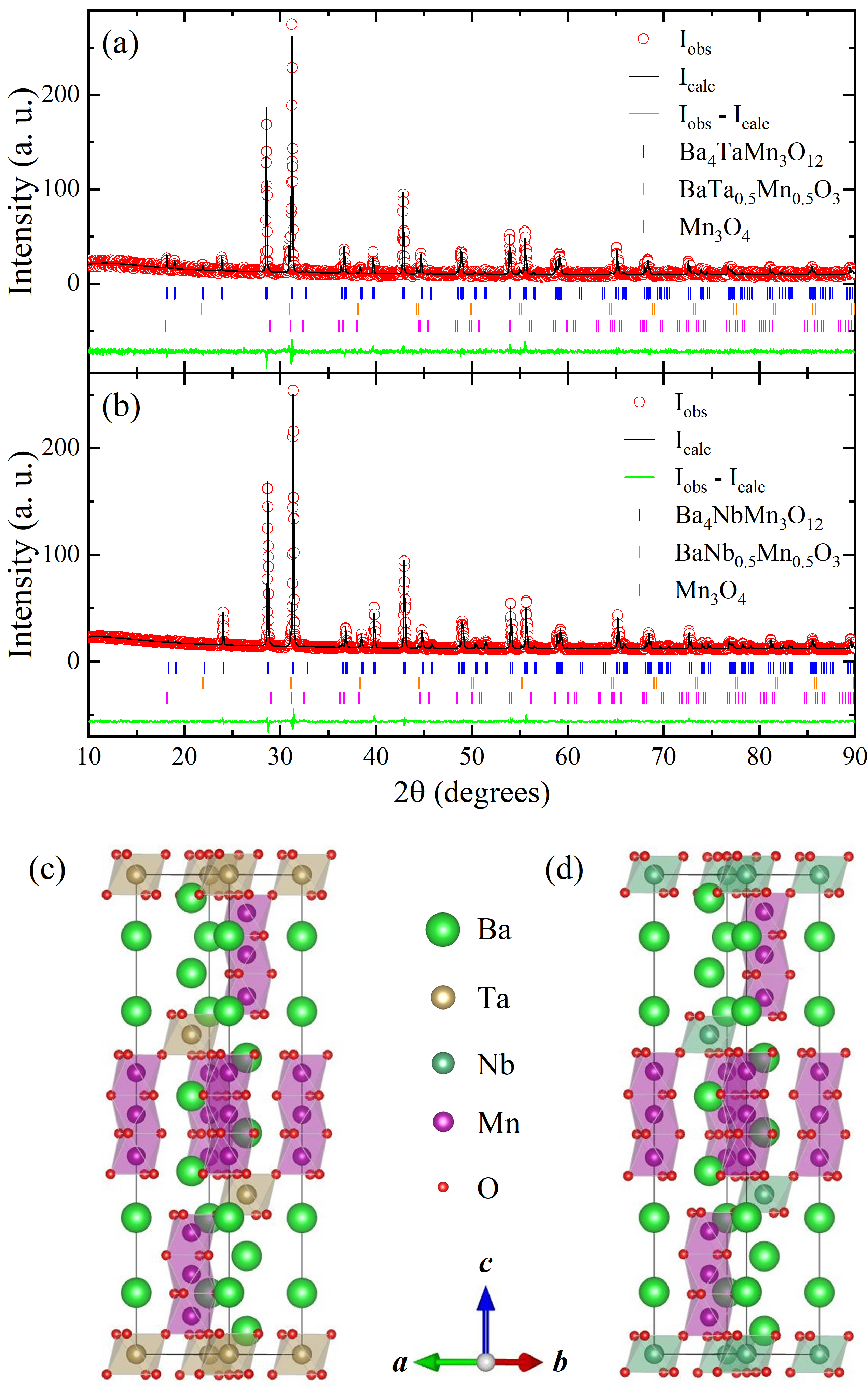}
\caption{(a), (b) Structural refinements using the powder x-ray diffraction data collected at room temperature from \BTMO and \BNMO, respectively. (c), (d) Refined crystal structures. The Mn trimer is indicated by shaded purple polyhedra along the crystallographic $c$ axis.
}
\label{XRD}	
\end{figure}

\subsection{Crystal structure}
\label{sec:str}
To determine the crystal structure of \BXMO compounds, powder x-ray diffraction was performed, revealing the high quality of the sample. Consequently, these batches were used for the subsequent characterization measurements presented in this paper. Figure~\ref{XRD} shows the x-ray diffraction patterns of \BTMO and \BNMO samples collected at room temperature. Rietveld refinements confirmed that both \BTMO and \BNMO crystallized in a trigonal crystal structure (\sg, No.~166), consistent with the  previous research~\cite{Nguyen2019} on \BNMO. A minor secondary phase [$\sim$3\pct of disordered perovskite \BXMOSP] was identified in both compounds, which was also reported previously for \BNMO~\cite{Nguyen2019}. Figures~\ref{XRD}(c) and \ref{XRD}(d) illustrate the refined crystal structures, where three MnO$_{6}$ octahedra are connected by their faces and form a Mn$_{3}$O$_{12}$ trimer. The Mn trimers are alternatively connected by corner-sharing nonmagnetic TaO$_{6}$ or NbO$_{6}$ octahedra by forming 12 hexagonal layers in total within the unit cell. The extracted structural parameters of \BTMO and \BNMO are provided in Table~\ref{T_1} and Table~\ref{T_2}, respectively. Note that we found only a few percentages of the site mixing between Mn and Ta (Nb) sites.

\begin{table}[]
		\caption{Refined structural parameters for \BTMO at 300~K. Space group \sg, No.~166, a~=~5.72543(12) \AA, c~=~28.14875(87), $U$$_{\rm iso}$ (fixed)~=~0.001~\AA$^{2}$, goodness of fit (GOF)~=~1.08, R$_{p}$~=~7.33\,\pct, wR$_{p}$~=~9.25\,\pct. Mass fraction of \BTMOP~=~4.69(16)\,\pct and Mn$_3$O$_4$~=~2.6(5)\,\pct. Sites are assumed to be fully occupied.
	}
	\label{T_1}
	\setlength\extrarowheight{5pt}
	\setlength{\tabcolsep}{10pt}
	\begin{tabular}{ccccc}
		\hline
		Atom & Site &    x     &    y     &     z      \\ \hline
		Ba1  &  6c  &    0     &    0     & 0.1282(3)  \\
		Ba2  &  6c  &    0     &    0     & 0.2848(2)  \\
		Ta1  &  3a  &    0     &    0     &     0      \\
		Mn1  &  3b  &    0     &    0     &    0.5     \\
		Mn2  &  6c  &    0     &    0     & 0.4131(6)  \\
		 O1  & 18h  & 0.486(2) & 0.514(2) & 0.1270(9)  \\
		 O2  & 18h  & 0.489(3) & 0.511(3) & 0.2912(11) \\ \hline
	\end{tabular}
\end{table}

\begin{table}[]
	\caption{Refined structural parameters for \BNMO at 300~K. Space group \sg, No.~166, a~=~5.72996(4) \AA, c~=~28.16229(27), U$_{\rm iso}$ (fixed)~=~0.001~\AA$^{2}$, goodness of fit (GOF)~=~1.27, R$_{p}$~=~2.55\,\pct, wR$_{p}$~=~3.27\,\pct. Mass fraction of \BNMOP~=~2.17(6)\,\pct and Mn$_3$O$_4$~=~2.00(18)\,\pct. Sites are assumed to be fully occupied.
	}
	\label{T_2}
	\setlength\extrarowheight{5pt}
	\setlength{\tabcolsep}{10pt}
	\begin{tabular}{ccccc}
		\hline
		Atom & Site &    x     &    y     &     z     \\ \hline
		Ba1  &  6c  &    0     &    0     & 0.1292(1) \\
		Ba2  &  6c  &    0     &    0     & 0.2846(1) \\
		 Nb  &  3a  &    0     &    0     &     0     \\
		Mn1  &  3b  &    0     &    0     &    0.5    \\
		Mn2  &  6c  &    0     &    0     & 0.4103(2) \\
		 O1  & 18h  & 0.489(1) & 0.511(1) & 0.1252(3) \\
		 O2  & 18h  & 0.490(1) & 0.510(1) & 0.2953(4) \\ \hline
	\end{tabular}
\end{table}

\subsection{Magnetic susceptibility}
\label{sec:Chi}

\begin{figure*} [t]
\includegraphics[width=0.49\linewidth]{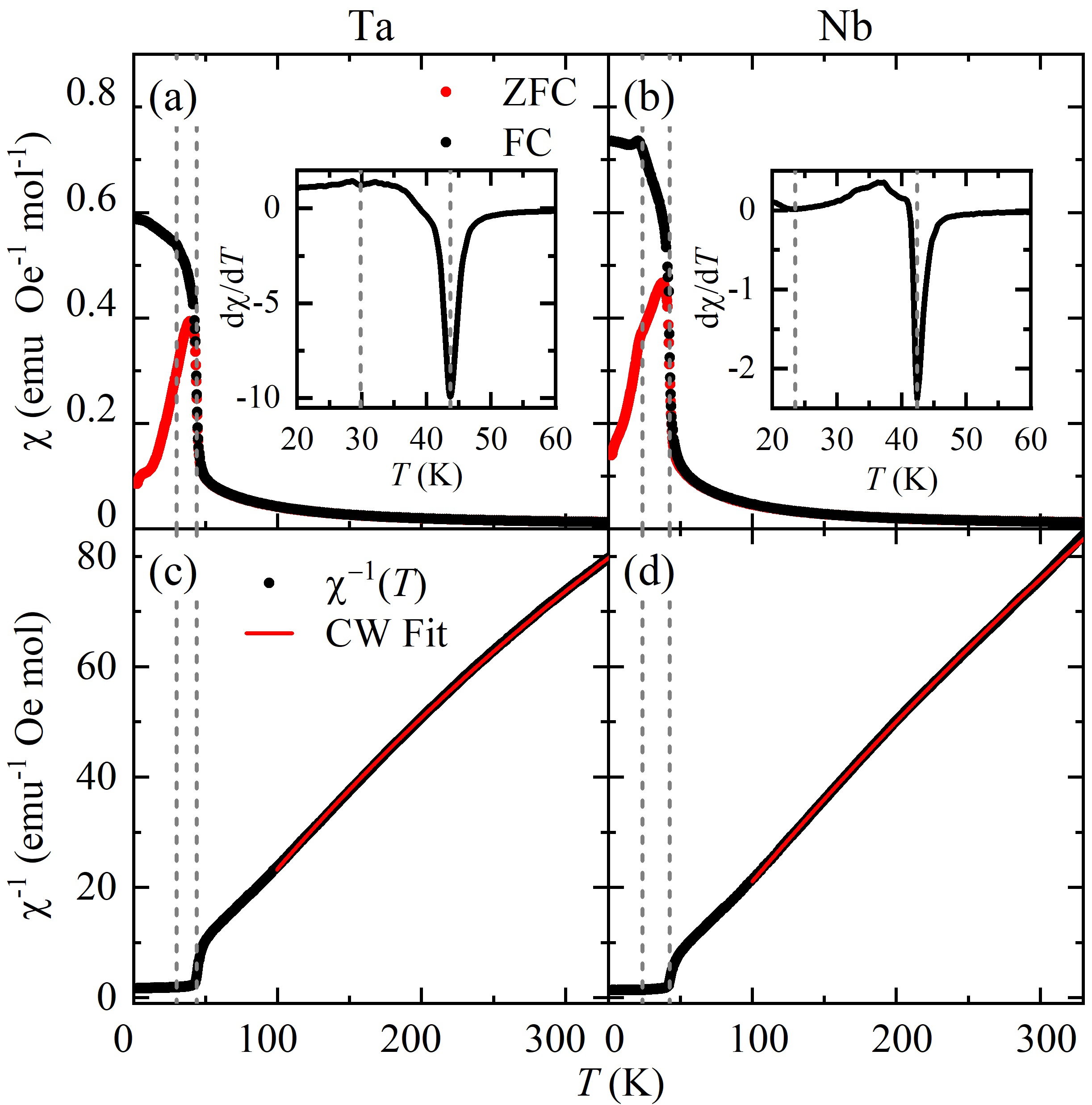}
\includegraphics[width=0.49\linewidth]{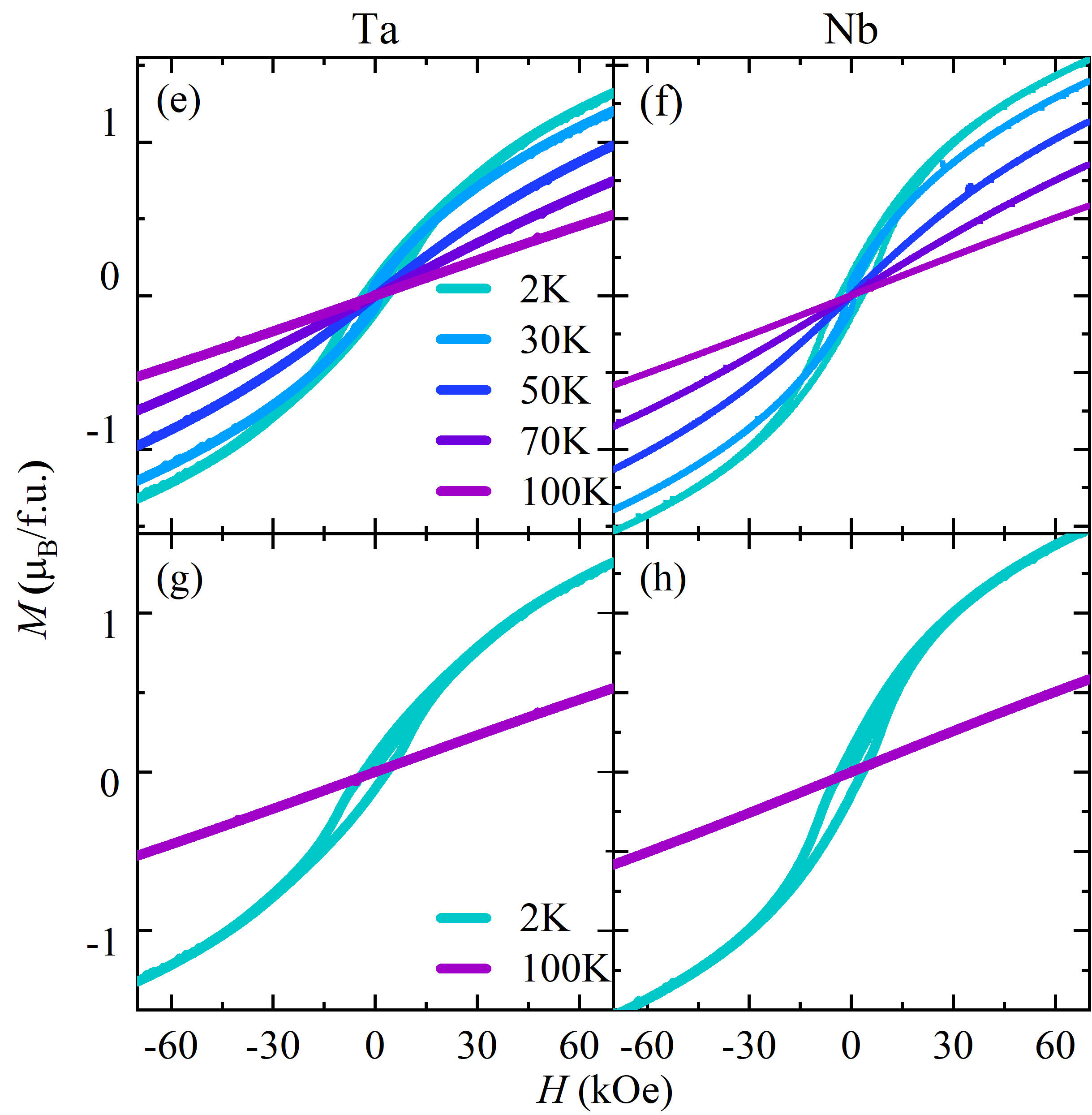}
\caption{(a), (b) Magnetic susceptibility collected at $H$ = 0.1~T. Insets: the first derivative of magnetic susceptibility revealed anomalous temperatures at 43.72~K (42.42~K) for $T_1$ and 29.77~K (23.5~K) for $T_2$ for \BTMO (\BNMO), respectively. (c), (d) Inverse magnetic susceptibilities (solid black lines) and Curie-Weiss (CW) fits (solid red lines). The gray dashed vertical lines illustrate the anomalous temperatures consistently found in (a)--(d). (e), (f) Magnetization ($M$) with magnetic field ($H$) of \BTMO and \BNMO measured at various temperatures ($T$) between $-$70 and 70~kOe. (g), (h) $M (H)$ data at 2 and 100~K.
}
\label{MTH}
\end{figure*}

\begin{table}
\begin{center}
\caption{Parameters obtained from CW fits using the inverse magnetic susceptibility $\rchi^{-1}(T)$ of \BTMO. $\theta$ is the CW temperature. $C$ (emu~K~Oe$^{-1}$~mol$^{-1}$) is the Curie constant. $\rchi_{0}$ (emu~Oe$^{-1}$~mol$^{-1}$) is the temperature-independent magnetic susceptibility. $\mu_\textrm{eff}$ is the effective magnetic moment.
}
\vspace{0.0cm}
\label{CW}
\setlength\extrarowheight{10pt}
\setlength{\tabcolsep}{6pt}

\begin{tabular}{c c c c c c c c}
	\hline
	   Range (K) & $\rchi_{0}$($\times$10$^{-3}$) & $\theta$ (K) & $C$       & $\mu_\textrm{eff}$ ($\mu_\textrm{B}$) \\ \hline
	   370~-~300    & 3.24                           & 23.14        & 2.86027 & 4.78                                  \\
	   370~-~200    & 3.73                           & 40.14        & 2.55749 & 4.52                                  \\
	   370~-~100    & 3.14                           & 28.78        & 2.83327 & 4.76                                  \\
	   275~-~100    & 3.15                           & 31.61        & 2.78916 & 4.72                                  \\ \hline
\end{tabular}
\end{center}
\end{table}

\begin{table}
\begin{center}
\caption{Parameters obtained from CW fits using the inverse magnetic susceptibility $\rchi^{-1}(T)$ of \BNMO.
}
\vspace{0.0cm}
\label{CW_2}
\setlength\extrarowheight{10pt}
\setlength{\tabcolsep}{6pt}

\begin{tabular}{c c c c c c c c}
	\hline
	Range (K)              & $\rchi_{0}$($\times$10$^{-3}$) & $\theta$ (K) & $C$       & $\mu_\textrm{eff}$ ($\mu_\textrm{B}$) \\ \hline
	370~-~300                 & -5.21                          & -100.08      & 7.3661  & 7.67                                  \\
	370~-~200                 & -2.27                          & -33.45       & 5.14962 & 6.42                                  \\
	370~-~100                 & 0.73                           & 24.70        & 3.40414 & 5.22                                  \\
	275~-~100                 & 2.06                           & 34.71        & 2.96607 & 4.87                                  \\
	275~-~100~\cite{Nguyen2019} & 3.07            & $-$~4        & -             & 4.82                                  \\ \hline
\end{tabular}
\end{center}
\end{table}

\begin{figure*} [t]
\includegraphics[width=0.49\linewidth]{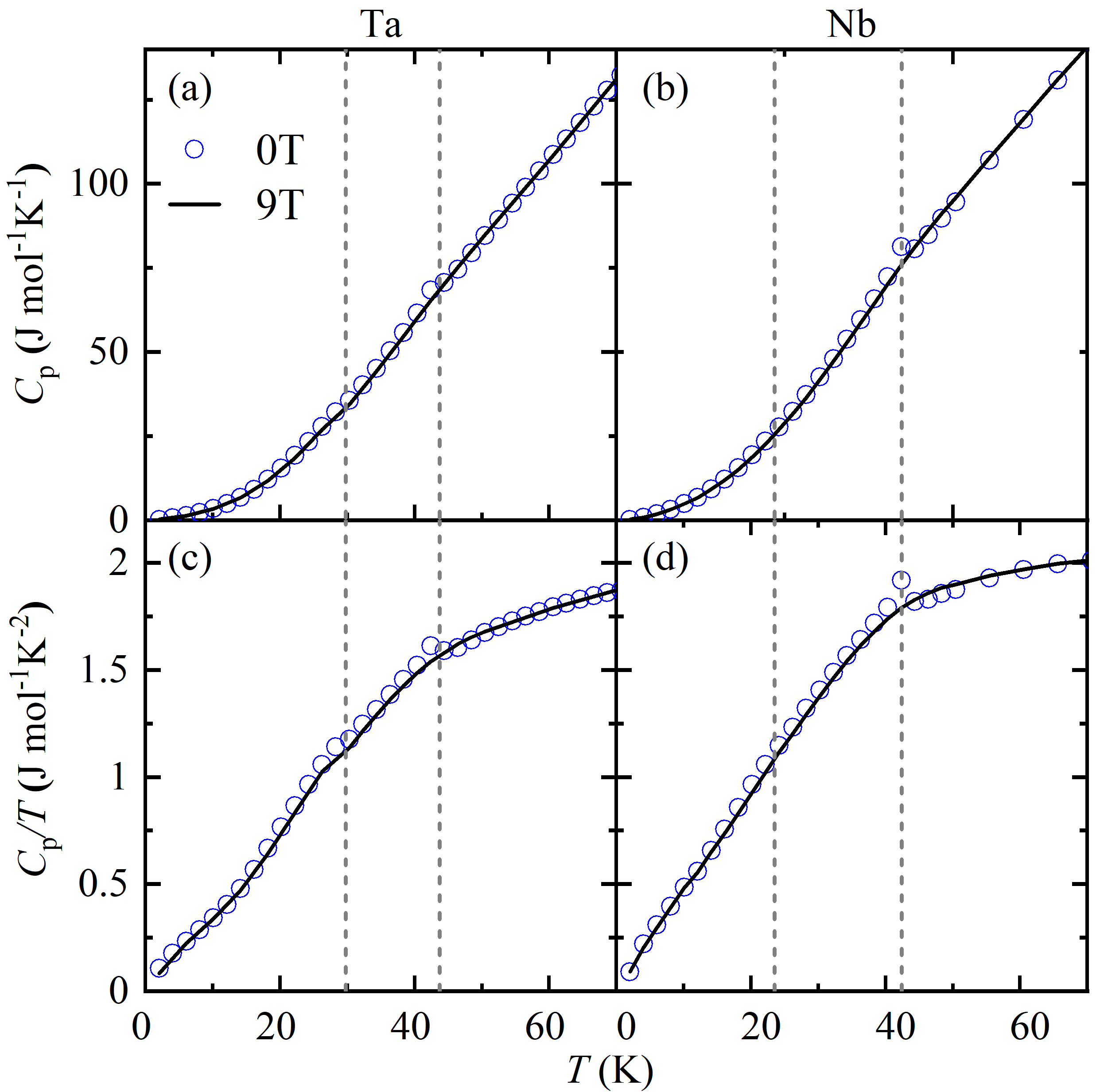}
\includegraphics[width=0.49\linewidth]{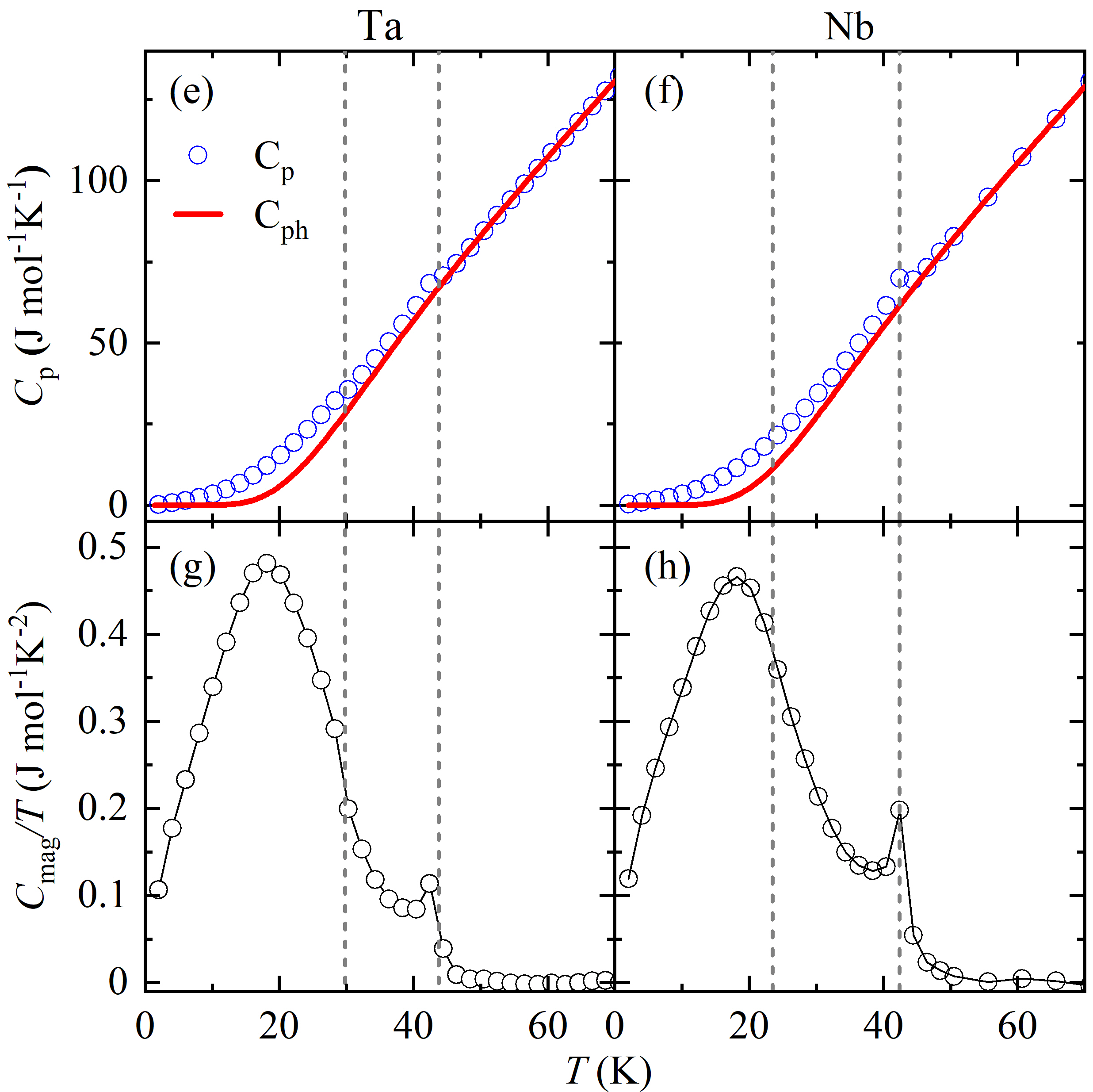}
\caption{Heat capacity ($C_{\rm p}$) data and analysis. (a), (b) $C_{\rm p}$ at 0 and 9~T between 2 and 160~K. (c), (d) Heat capacity divided by temperature, $C_{\rm p}$/$T$. Two lambda-like anomalous temperatures are noted by gray dashed vertical lines in (a)--(d); 43.72~K (42.42~K) for \BTMO (\BNMO) and 29.77~K (23.5~K) for \BTMO and (\BNMO). (e), (f) $C_{\rm p}$ overplotted with estimated phononic contributions (red lines; see texts). (g), (h) Magnetic contribution to heat capacity divided by temperature ($C_\textrm{ph}$).
}
\label{HC_1}	
\end{figure*}

\begin{figure} [t]
\includegraphics[width=\linewidth]{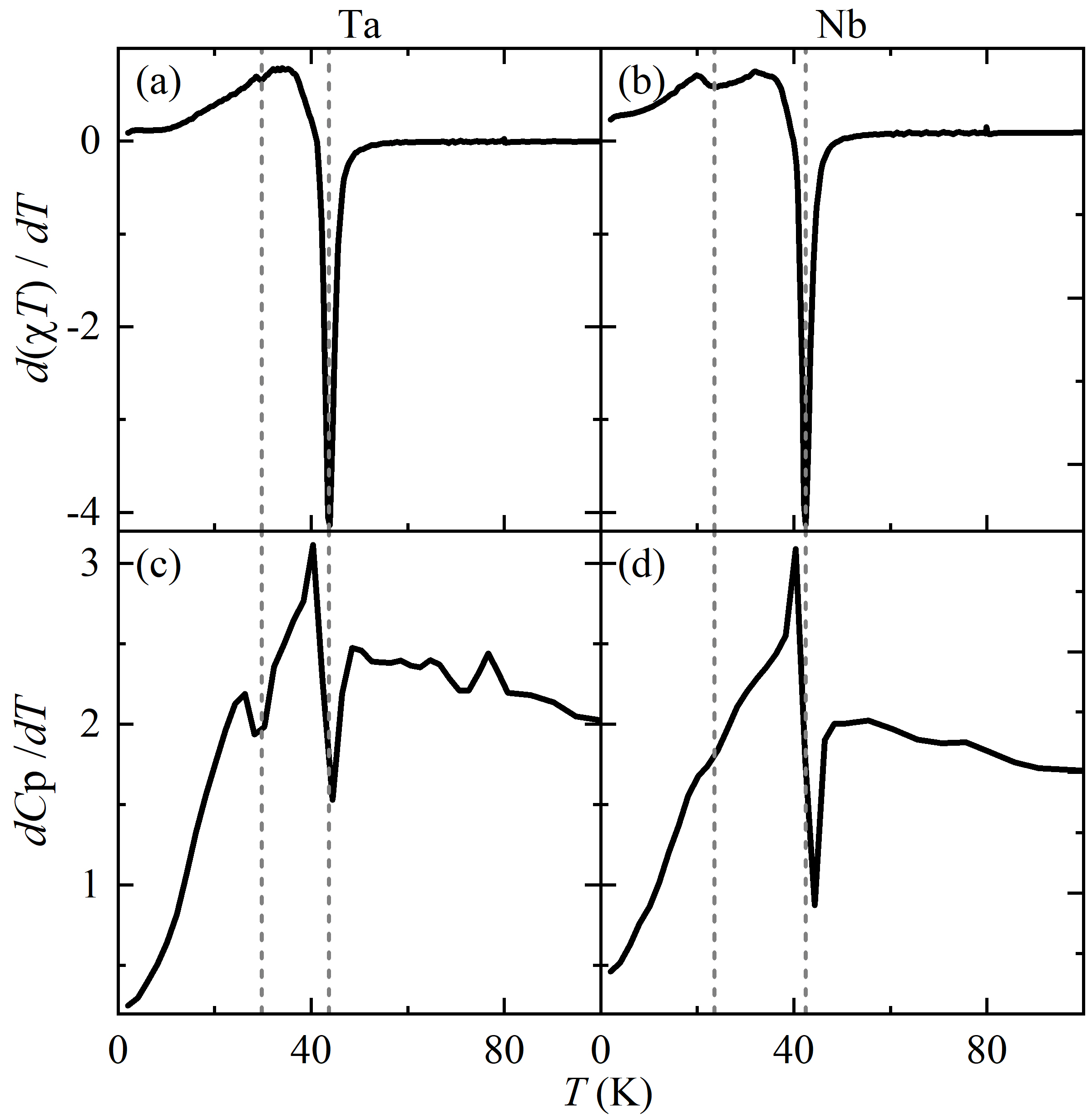}
\caption{Comparison of anomalous temperatures from (a), (b) $d$($\rchi$$T$)/$dT$ and (c), (d) $d$$C_{\rm p}$/$dT$ marked by gray dashed vertical lines. $T_{1}$~$\sim$~43.72~K and $T_{2}$~$\sim$~29.77~K from \BTMO and $T_{1}$~$\sim$~42.42~K and $T_{2}$~$\sim$~23.5~K from \BNMO.
}
\label{HC_2}	
\end{figure}

Magnetic susceptibility and magnetization measurements were then performed to understand the magnetism of \BTMO and \BNMO. Figures~\ref{MTH}(a) and \ref{MTH}(b) show the temperature-dependent ZFC and FC magnetic susceptibility data [$\rchi(T) = M/H$, where $M$ is the magnetization and $H$ is the applied magnetic field] measured at $H$ = 0.1 T from 300~K to 2~K, whereas Figs.~\ref{MTH}(c) and \ref{MTH}(d) exhibit their inverse magnetic susceptibilities. Magnetic susceptibility data demonstrated Curie-Weiss-like behaviors at high temperatures, but $\rchi(T)$ was sharply increased below 43.72~K and 42.42~K for \BTMO and \BNMO, respectively, consistent with the anomaly found in $d\rchi(T)/dT$ [insets of Figs.~\ref{MTH}(a) and \ref{MTH}(b)]. Below those temperatures, labeled as $T_{1}$, we observed bifurcations in the ZFC and FC data, consistent with a hysteresis loop that was clearly seen in $M(H)$ [Figs.~\ref{MTH}(e)-\ref{MTH}(h)]. As shown in Figs.~\ref{MTH}(g) and \ref{MTH}(h), the magnetic moments at 2~K with 7~T are 1.32~$\mu_\textrm{B}$/f.u. for \BTMO and 1.53~$\mu_\textrm{B}$/f.u. for \BNMO, suggesting that more measurements would be needed at higher magnetic fields. We found additional anomalies at 29.77~K and 23.5~K for \BTMO and \BNMO, respectively, which were also visible in $d$($\rchi$$T$)/$dT$ (Fig.~\ref{HC_2}).

The $\rchi^{-1}(T)$ data between 100 and 370~K were fitted using the Curie-Weiss (CW) equation, $\rchi(T) = \rchi_{0} + C/(T-\theta)$, where $\rchi_{0}$ is temperature-independent susceptibility, $C$ is the Curie constant, and $\theta$ is the CW temperature, as summarized in Tables~\ref{CW} and \ref{CW_2}. The effective magnetic moment was calculated using $\mu_\textrm{eff}$ = $\sqrt{3k_{B}C/N_{A}}$~\cite{Ali2016}, where $N_{A}$ the Avogadro constant and $k_{B}$ is the Boltzmann constant. Our fitting results in the temperature range between 100 and 275~K showed that $\mu_\textrm{eff}$~=~4.72 (4.87) $\mu_\textrm{B}$ for \BTMO (\BNMO), which is consistent with both a previous report on \BNMO~\cite{Nguyen2019} and the expected 4.9~$\mu_\textrm{B}$ from two magnetic structural candidates with the $S=2$ trimer, which was proposed for \BNMO~\cite{Nguyen2019}. Also, a positive temperature-independent susceptibility $\rchi_{0}~=~3.15\times10^{-3}~(2.06\times10^{-3}$)~emu~Oe$^{-1}$~mol$^{-1}$ for \BTMO (\BNMO) is similar to that reported in \BNMO~\cite{Nguyen2019}, reminiscent of the van Vleck paramagnetism~\cite{Mugiraneza2022}.

These two observations are compatible with an idea of locally-aligned magnetic moments even at high temperatures with remnant susceptibility~\cite{Nguyen2019} after the local moments are established within the trimer. It could be contributed from the mixing between the ground and excited states~\cite{Carlin1986} within the Mn trimer in \BXMOshort. Thus, all moments are not freely fluctuating, but the local antiferromagnetic moment within the trimer is already formed even at room temperature, which will not contribute to the Curie-Weiss temperature. This means that the positive Curie temperature reflects the subdominant ferromagnetic exchange interaction. In this view, the extracted parameters can be interpreted with the trimer unity rather than individual spins of Mn. This picture is also consistent with the localized part ($S=3/2$) of the extracted effective moment. The delocalized part ($S=1/2$) will be discussed soon. In \BXMOshort, the Curie-Weiss temperature is positive, which indicates the effective ferromagnetic intertrimer coupling. We note that positive values of $\rchi_{0}$ are also observed in other related compounds, such as \BNMO~\cite{Nguyen2019}, Ba$_{4}$NbRu$_{3}$O$_{12}$~\cite{Nguyen2018}, and Ba$_{4}$Nb$_{0.8}$Ir$_{3.2}$O$_{12}$~\cite{Thakur2020}. We also tried the Curie-Weiss fit without the $\rchi_{0}$ term (not shown), but this gave completely nonphysical results, such as wrong effective magnetic moments. Thus, the sizable temperature-independent paramagnetic term is essential for reliable fits to extract the physically sensible parameters for \BXMOshort.

For a more systematic analysis, we also performed the CW fits with a different range of temperatures, as summarized in Table~\ref{CW} and Table~\ref{CW_2} for \BTMO and \BNMO, respectively. The effective magnetic moments are comparable and $\theta$ values fluctuate in all trials in \BTMO. This means that the antiferromagnetically coupled moments within the trimer are robust up to 370~K for \BTMO by giving the compatible, effective magnetic moment with that of \BNMO from Ref.~\cite{Nguyen2019}.

On the other hand, the reported CW parameters in the literature~\cite{Nguyen2019} are somewhat different from our parameters from \BNMO, as compared in Table~\ref{CW_2}. The effective moments extracted from fits in \BNMO show an increasing trend as the lower bound of the fitted temperature becomes higher. Also, the CW temperatures become more negative simultaneously, consistent with the trend of $\rchi_{0}$ values. These behaviors of $\rchi_{0}$ and $\mu_\textrm{eff}$ signal that the three Mn spins behave more independently towards higher temperatures in \BNMO, unlike \BTMO. The fitting parameters $\rchi_{0}$ and $\mu_\textrm{eff}$ for both \BXMOshort are consistent with results from Ref.~\cite{Nguyen2019} when the fitting was done in the identical temperature range (between 100 and 275~K). Interestingly, the effective magnetic moment extracted from \BNMO at the higher temperature range is comparable to the values expected when all moments freely respond to the external magnetic field in the trimer unit having two Mn$^{+4}$ ions and one Mn$^{+3}$ ion; for instance, 6.16 (for the low-spin model) or 7.35 (for the high-spin model) $\mu_B$/f.u~\cite{Nguyen2019}. As the temperature is lowered, the three Mn spins form the antiferromagnetic trimer even above the long-range magnetic ordering temperature and yield the effective magnetic moment, comparable to the $S=2$ system. These results suggest that the bonding of Mn spins in the trimer is stronger for \BTMO than \BNMO. Accordingly, a similar behavior for $\rchi_{0}$ and $\mu_\textrm{eff}$ is expected for \BTMO at higher temperatures above 370~K. We note that Ref.~\cite{Nguyen2019} proposed the ferrimagnetic transition behavior from $\rchi(T)$ with a negative $\theta$ value for \BNMO, which means the dominant antiferromagnetic intertrimer interaction. However, we did not observe such a trend in our experiments. $\theta$ values from our results are positive in both materials when using the same range of the fitted temperatures.

In Figs.~\ref{MTH}(e)-\ref{MTH}(h) the magnetization $M$ is plotted against the magnetic field $H$ (between $-$70 and 70~kOe) with temperatures. At 100~K, the $M(H)$ varies linearly and behaves similarly to a paramagnetic system. The $M(H)$ curves form a hysteresis loop at lower temperatures, compared in Figs.~\ref{MTH}(g)-\ref{MTH}(h), where a hysteresis loop in the $M(H)$ data at 2~K disappears at 100~K. This is consistent with the bifurcation anomaly seen in magnetic susceptibility in Figs.~\ref{MTH}(a) and ~\ref{MTH}(b).

\begin{figure} [t]
\includegraphics[width=\linewidth]{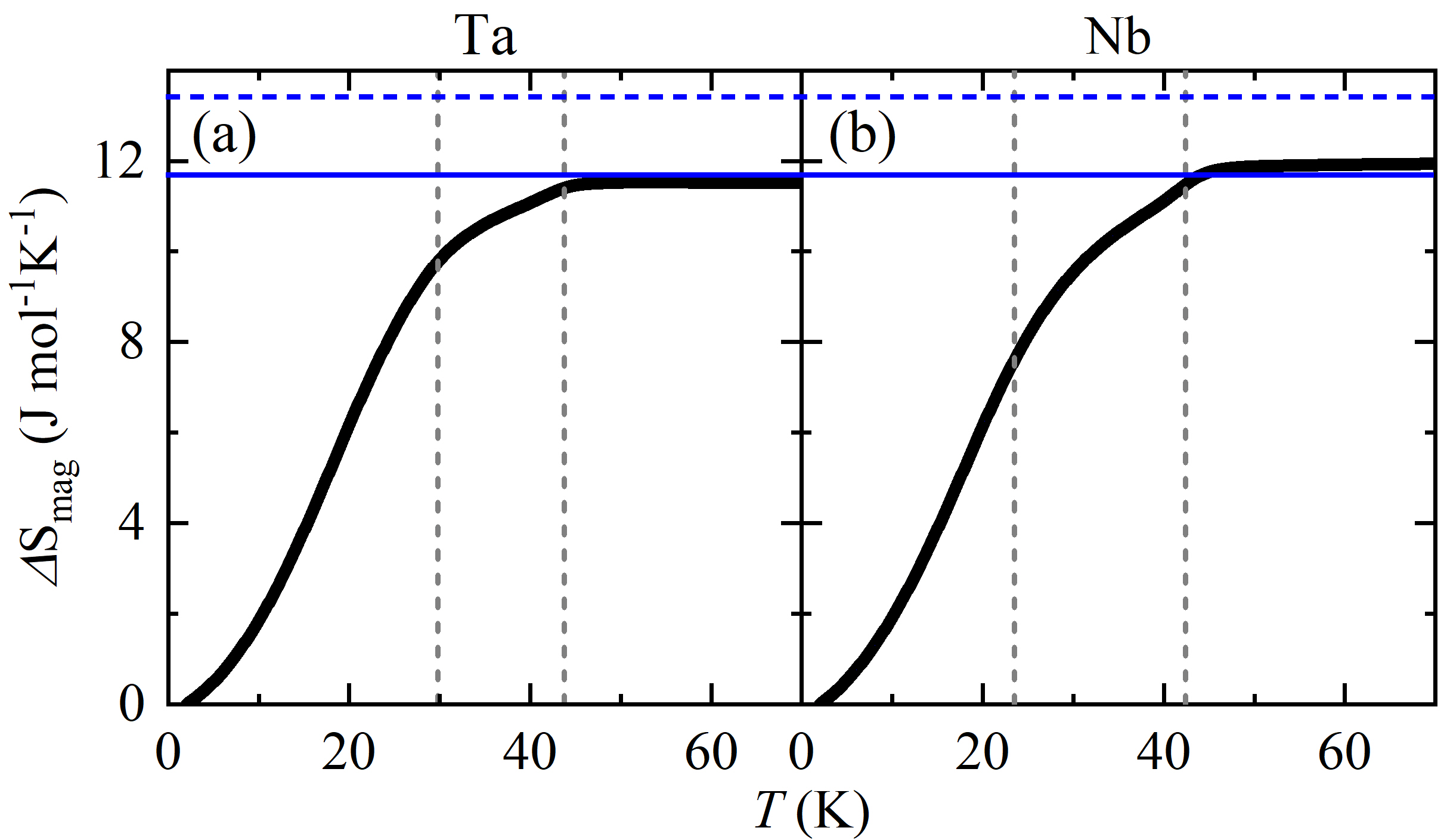}
\caption{Evolution of the magnetic entropy [$\Delta$$S$$_{\rm mag}$(T)] of (a) \BTMO and (b) \BNMO. The blue horizontal solid (dashed) lines present the expected entropy for the $S = 3/2$ ($S = 2$) trimer (see text).
}
\label{HC_4}	
\end{figure}

\subsection{Heat capacity}
\label{sec:Cp}
We performed heat capacity measurements on the same samples to understand the nature of the anomalous features found in the magnetic susceptibility data. Figures~\ref{HC_1}(a) and \ref{HC_1}(a) present the heat capacity (C$_{\rm p}$) of \BTMO and \BNMO, measured between 160 and 2~K at 0 and 9~T. The normalized heat capacity measurements (C$_{\rm p}$/$T$) of \BTMO and \BNMO are shown in Figs.~\ref{HC_1}(c) and \ref{HC_1}(d) as a function of temperature. Small lambda-like anomalies are observed around $T_{1}$ for both \BTMO and \BNMO at 0~T. As the temperature was lowered further, heat capacity exhibited additional anomalies at $T_{2}$~$\sim$~29.77~K for \BTMO and $T_{2}$~$\sim$~23.5~K for \BNMO. Both anomalies at $T_{1}$ and $T_{2}$ were also seen in $d$($\rchi$$T$)/$dT$ (Fig.~\ref{HC_2}). In the heat capacity data at 9~T, both anomalies are suppressed, indicating their magnetic origins.

Heat capacity data of magnetic insulators typically consist of a phononic (C$_\textrm{ph}$) and magnetic part (C$_\textrm{mag}$). A normal way to separate these parts is to subtract the heat capacity data of a nonmagnetic compound having the same crystal structure. However, in the absence of such a nonmagnetic analog, we fitted our heat capacity data with the Debye-Einstein model~\cite{Kittel2004, Caslin2014, Sebastian2021}, a phenomenological approach to capture the primary characteristic of the complex lattice dynamics. To estimate the lattice contribution, we used a combined model with one Debye term and two Einstein terms, as given by
\begin{equation}
C_{\rm ph} (T) = f_{D}C_{D}(\theta_{D}, T) + \sum_{i=1}^{2} g_{i}C_{E_i}(\theta_{E_i}, T).
\label{D_E}
\end{equation}

The first term in Eq.~(\ref{D_E}) is the Debye term (responsible for the acoustic modes), which is given by
\begin{equation}
C_{D}(\theta_{D}, T) = 9nR\left({\frac{T}{\theta_{D}}}\right)^{3} \int_{0}^{\theta_{D}/T} \frac{x^4e^x}{(e^x-1)^2} dx,
\label{Debye}
\end{equation}
where $n$ is the number of moles, $R$ is the universal gas constant, $\theta_{D}$ is the characteristic Debye temperature, and $x = \frac{\hbar \omega}{k_{B}T}$, where $\omega$ is the vibrational frequency. The second term contains two Einstein terms (optical modes) and is expressed as
\begin{equation}
C_{E_{i}}(\theta_{E_{i}}, T) = 3nR\left( \frac{\theta_{E_{i}}}{T} \right)^2 \frac{e^{\theta_{E_{i}}/T}}{(e^{\theta_{E_{i}}/T}-1)^2},
\label{Eienstien}
\end{equation}
where $\theta_{E_{i}}$ is the characteristic Einstein temperature. Since one formula unit of \BXMOshort consists of 20 atoms, the summed number of $f_{D}$ and $g_{i}$ in Eq.~(\ref{D_E}) is 20, which follows Debye and Einstein statistics, respectively. The heat capacity data, C$_{\rm p}$($T$), between 50 and 150~K are fitted well with Eq.~(\ref{D_E}) and the obtained parameters were $f_{D}$~$\simeq$~6, $g_{1}$~$\simeq$~8, $g_{2}$~$\simeq$~6, $\theta_{D}$~$\simeq$~1016 (710)~K, $\theta_{E_{1}}$~$\simeq$~404 (285)~K, and $\theta_{E_{2}}$~$\simeq$~145 (110)~K for \BTMO (\BNMO) [red solid lines in Figs.~\ref{HC_1}(e) and \ref{HC_1}(f)]. We also tested the fit using a Debye term and a Debye plus Einstein term, but the fits did not work. Thus we used a Debye term and two Einstein terms as a minimal set of terms in the representative fit in this work. We estimated the magnetic contribution to heat capacity ($C_\textrm{mag}$) by subtracting the phonon contribution ($C_\textrm{ph}$) from the measured heat capacity ($C_{\rm p}$) and we extrapolated the fitted curve for temperatures between 160 and 2~K [see solid red lines in Figs.~\ref{HC_1}(e) and ~\ref{HC_1}(f)]. Figures~\ref{HC_1}(g) and \ref{HC_1}(h) show the values of $C_\textrm{mag}$/$T$ between 2 and 70~K, which reveal a sharp peak around $T_{1}$ and a broad hump centered at approximately 20~K in both compounds, capturing the magnetic contribution. Note that the number of moles ($n$) in Eqs.~(\ref{Debye}) and (\ref{Eienstien}) was not used in our fits.

\begin{figure} [t]
\includegraphics[width=\linewidth]{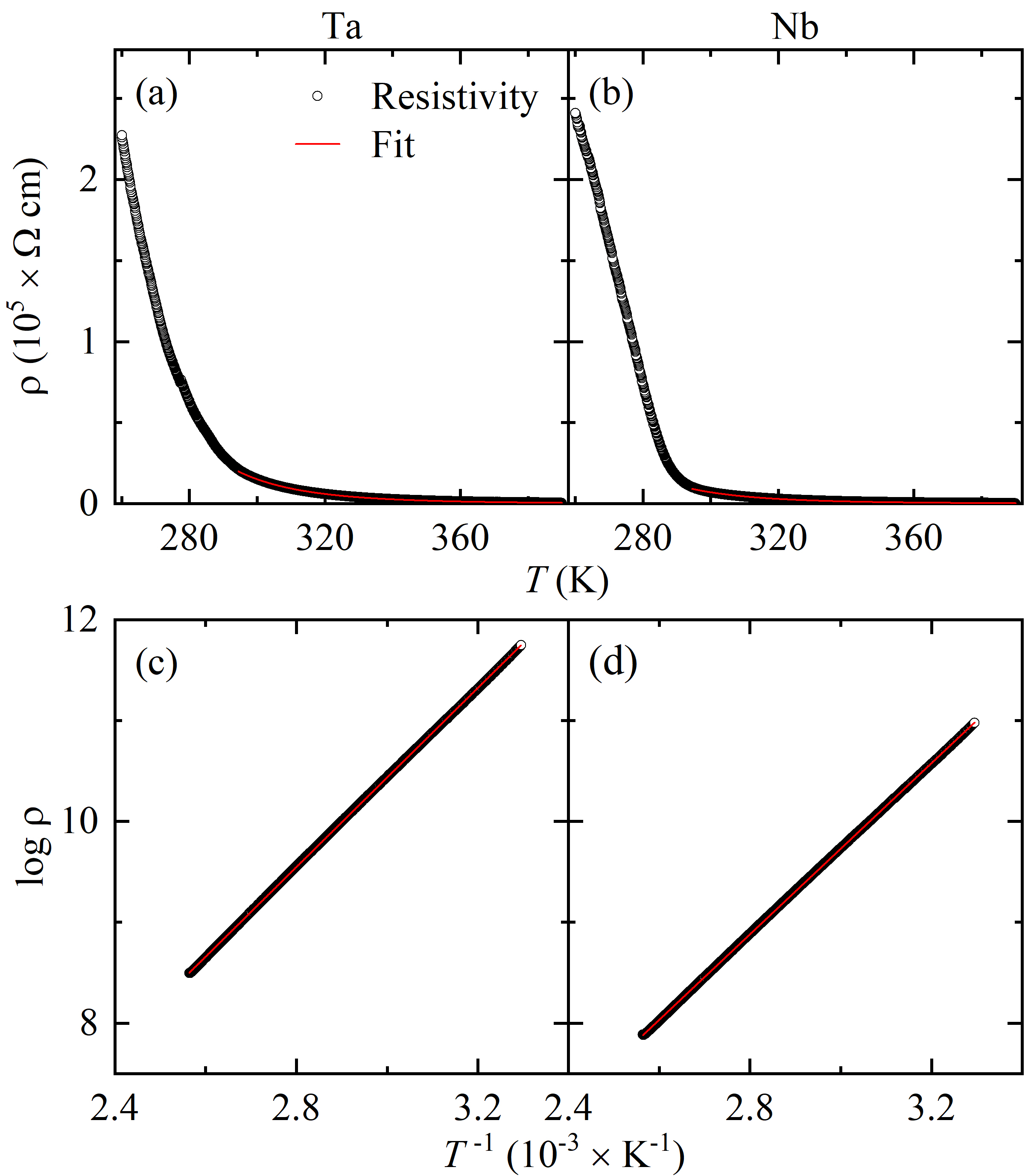}
\caption{(a), (b) Resistivity $\rho$($\Omega$ cm) versus temperature and (c), (d) log $\rho$ versus $T^{-1}$(K$^{-1}$) of \BTMO and \BNMO, respectively, measured on sintered rectangular pellets. Red solid lines represent the fitted curves (see text).
}
\label{Res}
\end{figure}

\begin{figure} [t]
\includegraphics[width=\linewidth]{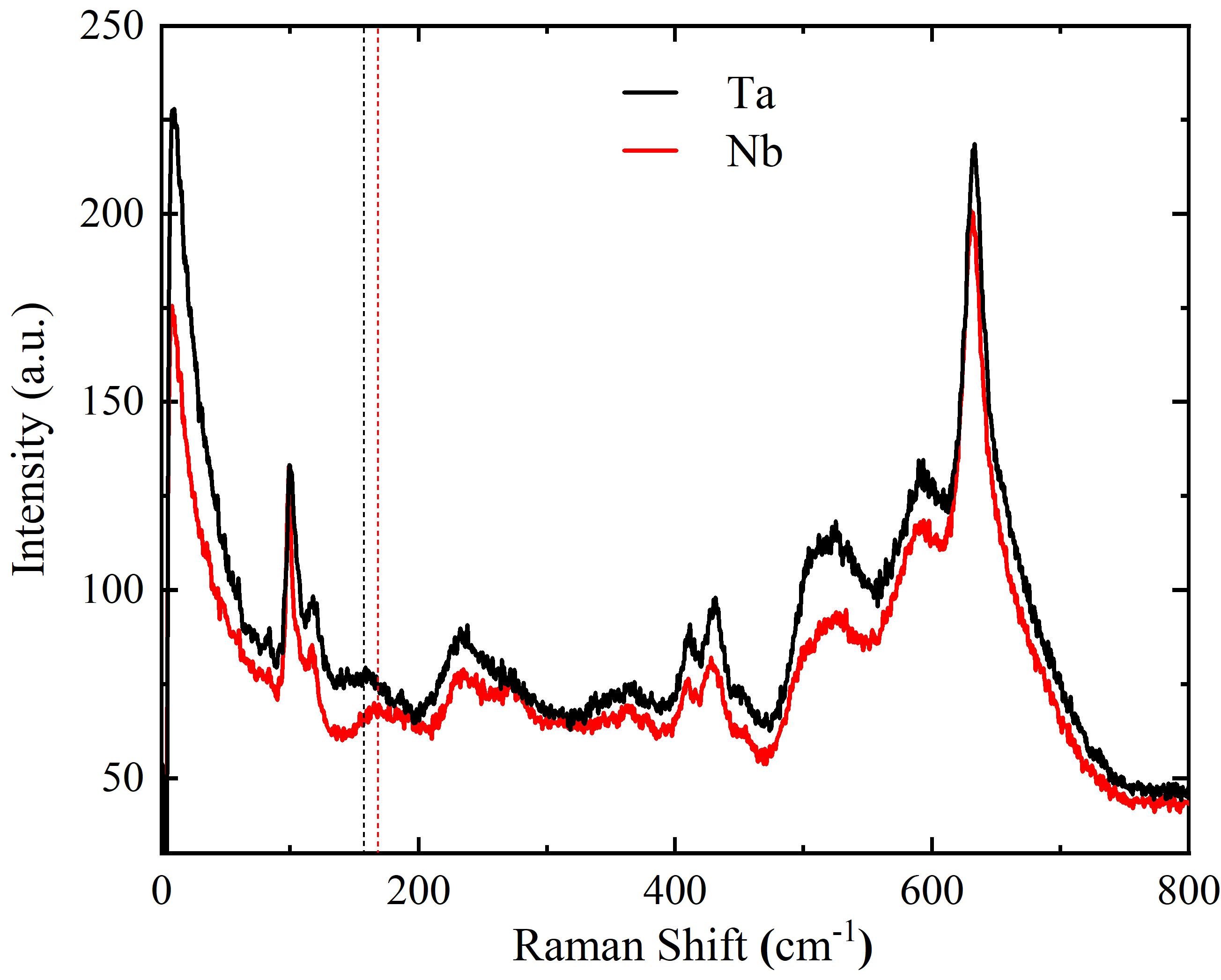}
\caption{Comparison of Raman spectra measured at room temperature.}
\label{Raman}	
\end{figure}

\begin{figure*} [t]
\includegraphics[width=\linewidth]{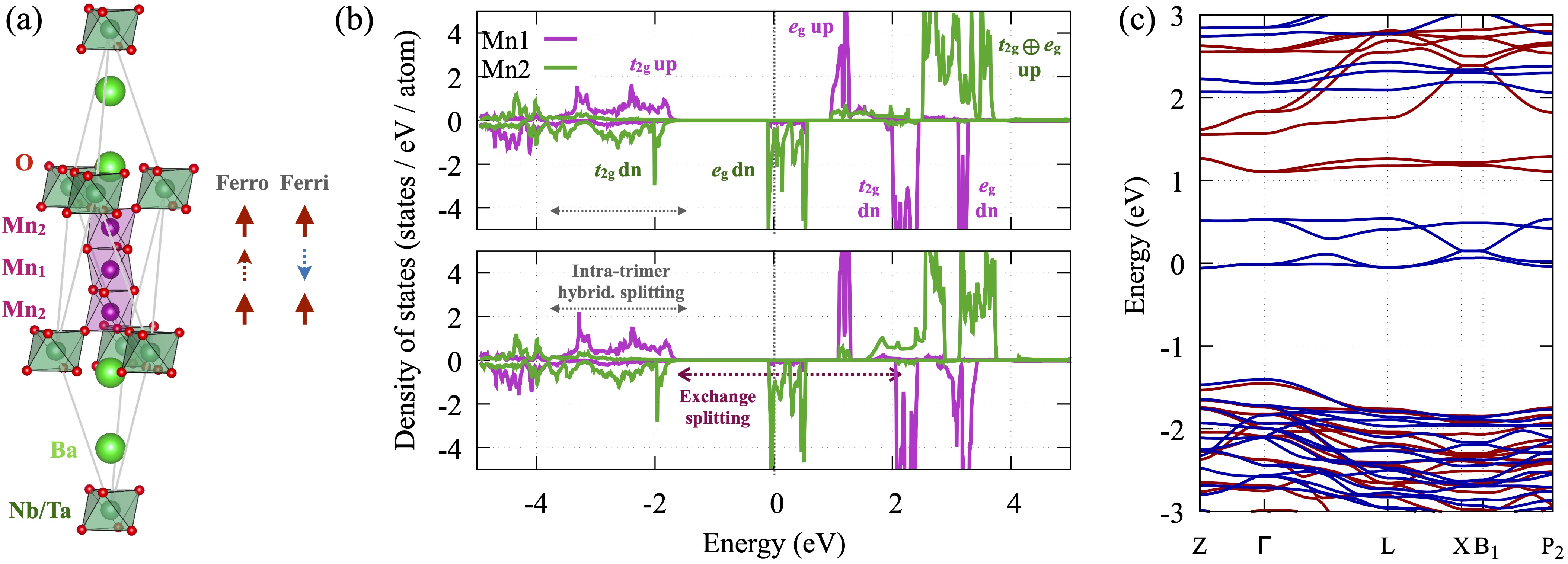}
\caption{(a) Primitive rhombohedral unit cell of \BNMO and \BTMO, where the two simplest magnetic configurations at Mn trimer are shown on the right. Note that Mn atoms at the center and two sides of the Mn trimer are denoted as Mn$_1$ and Mn$_2$, respectively (consistent with Tables~\ref{T_1} and~\ref{T_2}). $S=3/2$ moments at Mn$_1$ sites are depicted as dotted arrows. (b) Projected density of states (PDOS) for \BNMO (a top panel) and \BTMO (a bottom panel), where the energy splitting by hybridization and exchange energy is shown. (c) Band structure of \BTMO with the ferrimagnetic trimer in the primitive unit cell shown in (a). Red and blue lines depict spin-up and -down bands, respectively. Note that the band of \BNMO is qualitatively the same as that of \BTMO (not shown). Conventions for high-symmetry points follow the ones described in Ref.~\cite{Setyawan2010}.
}
\label{DFT_dos_band}
\end{figure*}

Figure~\ref{HC_4} exhibits the evolution of the magnetic entropy ($\Delta$$S$$_{\rm mag}$) between 2 and 70~K. The change in magnetic entropy was calculated by integrating C$_\textrm{mag}$/$T$ with respect to temperature between 2 and 70~K, which yielded the saturated value of $\Delta$$S$$_{\rm mag}$ = 11.52~J~mol$^{-1}$~K$^{-1}$ for \BTMO and $\Delta$$S$$_{\rm mag}$ = 11.94~J~mol$^{-1}$~K$^{-1}$ for \BNMO. Both values are smaller than the expected entropy for the $S = 2$ magnet (dashed horizontal lines in Fig.~\ref{HC_4}); $\Delta$$S$$_{\rm mag}~=~R$\rm{ln}$(5)$~$\sim$~13.38~J~mol$^{-1}$~K$^{-1}$, where $R$ is the gas constant. However, these values are nearly the same as that of the unexpected $S = 3/2$ antiferromagnetic trimer (horizontal solid lines in Fig.~\ref{HC_4}), $\Delta$$S$$_{\rm mag}~=~R$\rm{ln}$(4)$~$\sim$~11.53~J~mol$^{-1}$~K$^{-1}$. Thus our heat capacity analysis suggests the $S = 3/2$ trimer in both compounds, which could not be determined in a previous report~\cite{Nguyen2019}. With the uncertainty of the empirically estimated phonons, the agreement of the saturated moments with the expected one for the $S = 3/2$ trimer is satisfactory. The reduced magnetic entropy from that of the $S = 2$ trimer to the $S = 3/2$ trimer can be interpreted by one delocalized electron within the trimer in a partial molecular orbital state (to be explained in Sec.~\ref{sec:abinitio}).

To determine the anomalous temperature more systematically, we compared $d$($\rchi$$T$)/$dT$ and $d$$C_{\rm p}$/$dT$, finding that the extracted anomalous temperatures (gray dashed vertical lines) from both measurements agreed very well, as shown in Fig.~\ref{HC_2}. The additional anomaly ($T_{2}$) in the magnetic system could appear for various reasons, such as an additional magnetic transition, or a (continuous) spin reorientation~\cite{White1969}. It is also possible that a spin-glass transition~\cite{Young2012} from a finite disorder occurs at $T_{1}$, followed by the long-range magnetic order at $T_{2}$. Their origins can be examined via other complementary measurements, such as neutron diffraction, in future studies.

\subsection{Resistivity}
\label{sec:Resistivity}
Figure~\ref{Res} shows the resistivity data of \BTMO and \BNMO between 260 and 390~K. Data below 260~K were not obtained due to the exceedingly large resistivity that prevented any measurements in that region in our resistivity option of PPMS. Consequently, we fitted the data between 300 and 390~K with the resistivity equation $\rho$ = $\rho_o$ $e^{\frac{E_a}{k_B T}}$ to obtain activation energy of transport, $E_a$, which was found to be 0.383 eV for \BTMO and E$_a$ = 0.365 eV for \BNMO. Thus, it was confirmed that both materials exhibited semiconductor properties.

\subsection{Raman scattering}
\label{sec:Raman}
Raman spectroscopy is a suitable technique for examining phonons sensitive to crystal structure variations. We performed Raman measurements on both compounds at room temperature to confirm the sample quality and explore lattice dynamics. Figure~\ref{Raman} compares similar phonons of two compounds. Although we included measurements up to 1100~cm$^{-1}$, multiple phonon peaks are merely observed beyond 800~cm$^{-1}$. The sharp phonon peaks observed indicate the good qualities of our polycrystalline samples. A phonon of \BTMO at approximately 180~cm$^{-1}$ are slightly softened, possibly because Ta ions are heavier than Nb ions. In contrast, Raman phonon spectra at higher energies are more similar, which could be explained by the lighter ions Mn and O. Raman measurements at lower temperatures (i.e., lower than the anomalous temperatures obtained from susceptibility and heat capacity measurements) might be sensitive in detecting bond angle changes and distances in the magnetically ordered state by spin-lattice coupling~\cite{Casto2015}. These measurements may also provide information on magnetic ground states and exchange interaction with respect to molecular orbital states~\cite{Ye2018}.

\section{{\it Ab initio} calculations}
\label{sec:abinitio}
To understand the electronic structure and magnetism of both compounds, we performed {\it ab initio} density functional theory (DFT) calculations. We initially assessed the magnetic nature of the Mn trimer and Fig.~\ref{DFT_dos_band}(a) shows its two simplest magnetic configurations---a ferromagnetic and a ferrimagnetic order in the primitive unit cell. Note that the ferrimagnetic order in this section means the ferrimagnetic trimer in the primitive unit cell. We found that the total energy per formula unit of the ferrimagnetic order was lower than that of the ferromagnetic case for \BNMO and \BTMO by 62.7~meV and 55.9~meV, respectively. This could be attributed to the antiferromagnetic coupling between Mn-Mn in the face-sharing MnO$_{6}$ octahedra originating from a strong direct orbital overlap between Mn 3$d$ orbitals discussed in previous literature~\cite{Streltsov2018}.

\begin{figure*} [t]
\includegraphics[width=\linewidth]{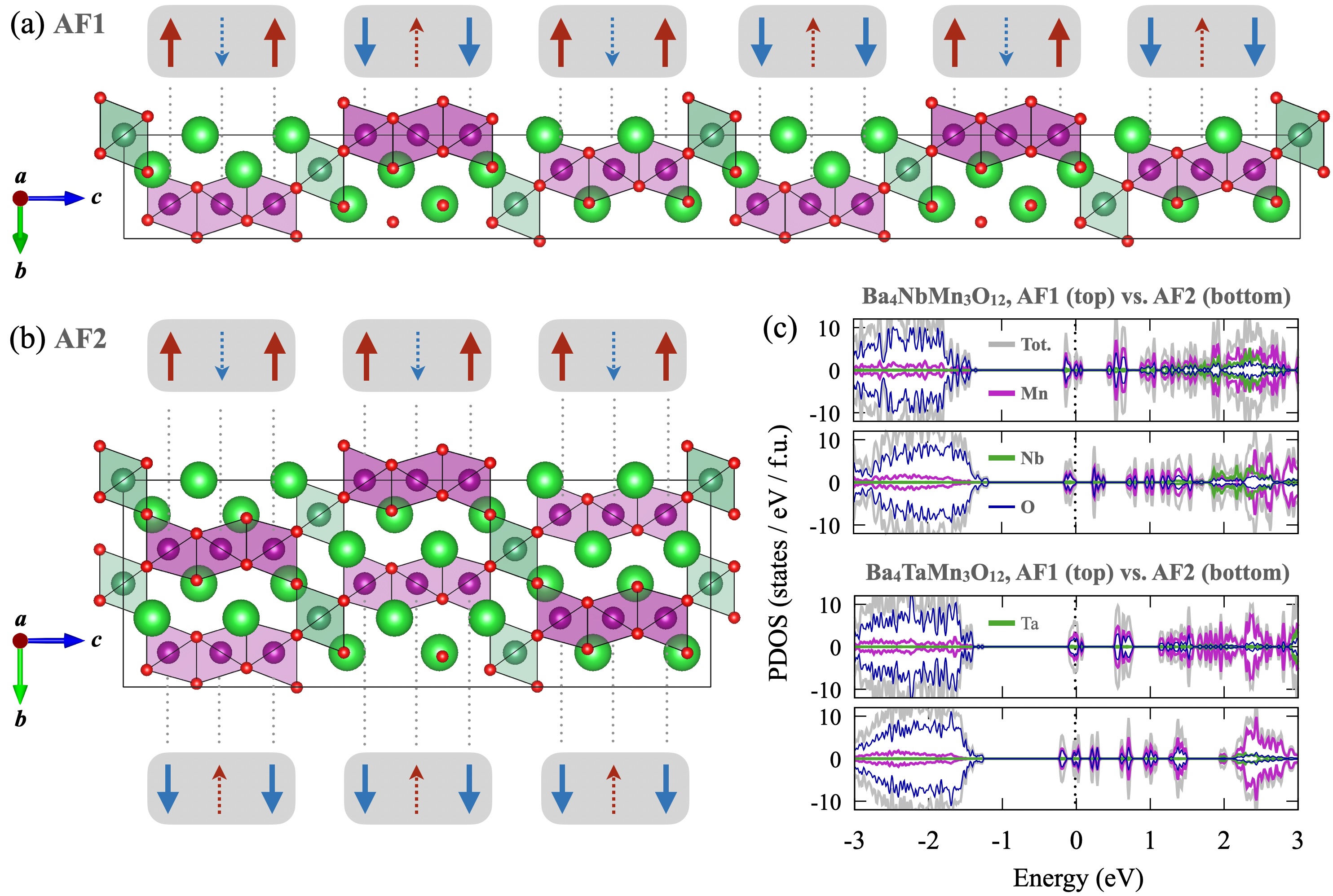}
\caption{Results of band structure calculations of the two simplest antiferromagnetic magnetic configurations. (a) AF1 and (b) AF2, which employ the Mn-trimer as a magnetic unit. (c) PDOS for \BNMO (upper panels) and \BTMO (lower panels), where the top and bottom ones correspond to PDOS of the AF1 and AF2 configuration, respectively.
}
\label{DFT_AF12}	
\end{figure*}

On the one hand, a molecular orbital (MO) state is proposed in a 4$d$~\cite{Streltsov2012} and 5$d$~\cite{Ye2018} transition-metal compound with a similar crystal structure (the trimer with the face-sharing octahedra) owing to the strong direct overlap of orbitals. On the other hand, we identified that the direct overlap between the face-sharing Mn $a_{\rm 1g}$ orbitals was significant but not strong enough to break the local moment scheme in both compounds. For instance, Fig.~\ref{DFT_dos_band}(b) presents the projected density of states (PDOS) of \BNMO (a top panel) and \BTMO (a bottom panel) in the ferrimagnetic configuration, where the atomic exchange splitting due to Hund's coupling is larger than the energy splitting induced by the direct orbital overlap in the face-sharing geometry. Hence the $t_{\rm 2g}$ shell was half filled with a robust $S=3/2$ local moment at each Mn site in the antiferromagnetic trimer. In Fig.~\ref{DFT_dos_band}(b), this is evident from six electrons occupying the $t_{\rm 2g}$ down-spin orbital and three electrons occupying the $t_{\rm 2g}$ up-spin orbitals.

Interestingly, unlike the $t_{\rm 2g}$ shell, one electron having the $e_{\rm g}$ character in an Mn$^{3+}$ ion was strikingly delocalized and preferred a metallic phase in the ferrimagnetic state. The $e_{\rm g}$ orbital at the central Mn site was almost empty, whereas one electron is equally shared by the $e_{\rm g}$ orbitals at the top and bottom Mn site, as illustrated in Fig.~\ref{DFT_dos_band}(b). This is reminiscent of a similar 4$d$ analog for \BRO~\cite{Klein2011, Streltsov2012}, where the middle ion of the Ru trimer was nonmagnetic. Hence, the total spin moment at the Mn trimer became $S=2$, which is decomposed into the localized $t_{2g}$ ($S=3/2$) and delocalized $e_{g}$ ($S=1/2$) components. Note that this observation was also consistent with our magnetic susceptibility data (Fig.~\ref{MTH}) and heat capacity data (Fig.~\ref{HC_4}).

Figure~\ref{DFT_dos_band}(c) shows the calculated band structure, revealing that the partially filled $e_{\rm g}$ band (solid blue lines) exhibits a strong two-dimensional character of an almost flat dispersion along with the Z-$\Gamma$ line parallel to the Mn trimer along the $c$ axis. In contrast, our resistivity data in Fig.~\ref{Res} strongly suggest that both \BNMO and \BTMO are gapped. This is due to a well known flaw of the DFT method, which cannot capture the paramagnetic Mott phase as it cannot describe fluctuating magnetic moments. However, the magnetic fluctuation diminishes in the long-range magnetic order below the transition temperature, where the DFT method can provide a qualitatively valid picture. The nature of the band gap can be examined via dynamical mean-field theory calculations in further study, by properly considering fluctuating magnetic moments.

To determine a possible magnetic ground state, we used the two simplest antiferromagnetic configurations (denoted as AF1 and AF2) by adopting the ferrimagnetic Mn trimer as the magnetic building block [see Figs.~\ref{DFT_AF12}(a) and \ref{DFT_AF12}(b)]. The AF1 state contains a magnetic unit cell that is twice enlarged along the $c$ axis compared to the parent nuclear unit cell with a wave vector, $q$~=~(0, 0, 1/2), whereas the AF2 state has a doubled unit cell along the $a$ axis in an orthorhombic magnetic unit cell based on the in-plane antiferromagnetic order, $q$~=~(1/2, 0, 0). We set up the AF1 state based on information provided in Ref.~\cite{Streltsov2018}, where the antiferromagnetic order within the trimer is coupled ferromagnetically in the $ab$ plane and antiferromagnetically along the $c$ axis. In our calculations, we detected that the total energy of the orthorhombic AF2 configuration is lower by 178.9~meV and 85.0~meV per formula unit than that of the AF1 state for \BNMO and \BTMO, respectively. We also found a similar tendency of the MO state in both compounds. Note that our calculated band gap of \BTMO is vanishingly small even in the orthorhombic configuration with $U_{\rm eff}$ = 4 eV for both compounds. We point out that the AF2 state is a collinear antiferromagnetic order, which can serve as a good approximate magnetic structure.

\section{Discussion}
\label{sec:discussion}
We are in a position to discuss the nature of the Mn trimer in \BXMO. Our first-principle calculations predict an unusual combination of localized and delocalized electrons; that is, the localized antiferromagnetic magnetic moments ($S=3/2$) within the trimer and one electron delocalized in the two-end Mn ions of the trimer. Our picture can be compatible with both magnetic susceptibility (Sec.~\ref{sec:Chi}) and heat capacity results (Sec.~\ref{sec:Cp}).

In magnetic susceptibility experiments, we found that the effective magnetic moments from a paramagnetic temperature indicate the $S=2$ trimer in both compounds. They are compatible with the effective magnetic moments reported in Ref.~\cite{Nguyen2019} for \BNMO. At first, our results might be counterintuitive because of the localized $S=3/2$ trimer in \BXMOshort. However, they can be explained by the additional contribution from one delocalized electron in the trimer. For instance, the effective magnetic moments above the transition temperature reported in weakly ferromagnetic itinerant metals, such as ZrZn$_{2}$~\cite{Moriya1984} and Sc$_{3}$In~\cite{Moriya1984}, are much bigger than those expected from the small ordered magnetic moment; i.e., in ZrZn$_{2}$, the saturated magnetic moment is only about 0.12~$\mu_{B}$~\cite{Moriya1984, Moriya1985}, whereas the effective magnetic moment obtained from the Curie-Weiss-like law is about 1.4~$\mu_{B}$~\cite{Lacheisserie2002, Moriya1985}, close to the effective magnetic moment of spin-1/2, 1.73~$\mu_{B}$. These experimental observations are understood by the interaction of spatially extended modes of spin fluctuations, giving the Curie-Weiss-like behavior with a large effective magnetic moment~\cite{Moriya1984, Moriya1985, Moriya1991}. Since the experimental observations in itinerant magnetic systems are consistent with our observations, a similar microscopic mechanism could be applied to the delocalized electron in \BXMOshort.

Possible origins of the weak hysteresis in the magnetization data could be associated with various reasons, such as the ferrimagnetic moments deviating from the collinear antiferromagnetic moments and a spin glass feature from the finite disorder. Further magnetic susceptibility measurements could reveal the nature of the magnetic ground state. For instance, such states could be studied via ac magnetic susceptibility measurements, by probing domain walls of ferrimagnetic order, or by confirming freezing temperatures with the shift of the ac data by different oscillating frequencies~\cite{Topping2019}. These measurements will reveal the existence of finite disorder, which might result in the weak hysteric behavior in the magnetization data [Figs.~\ref{MTH}(g) and \ref{MTH}(h)].

In addition, we checked the possible existence of the ``unpaired'' $S=2$ moment in both compounds using the susceptibility data [Figs.~\ref{MTH}(a) and \ref{MTH}(b)]. If one $e_{g}$ electron is localized in one of the two-end Mn ions of the trimer without forming a partial molecular orbital, the fully localized $S=2$ moment will be possible in the sample. Since the unpaired $S=2$ will not have a presumed long-ranged spin-spin correlation with other normal trimers, we tabulated it with the paramagnetic term of $C_{\rm u}/T$ in the revised Curie-Weiss fit, where $C_{\rm u}$ means the Curie constant of the unpaired moment. The revised fits did not work and gave diverging results, meaning no significant fraction of the impurity state. This outcome is consistent with our additional DFT + $U_{\rm eff}$ calculations (not shown), which confirm that such a state is unstable in the realistic range of $U_{\rm eff}$ (3 $<$ $U_{\rm eff}$ $<$ 5~eV). Instead, we found that all initial configurations that started with the unpaired state converged to the partial molecular orbital state in our calculations.

In heat capacity measurements, we clearly showed that the magnetic entropy between 2 and 70~K is only $S=3/2$, not $S=2$ (Fig.~\ref{HC_4}) in both compounds. One might suggest that the reduction of the magnetic entropy in Fig.~\ref{HC_4} is led by the strong quantum fluctuations due to the frustrated magnetic exchanges~\cite{Zhong2019}. However, this seems less likely as the frustration indexes ($f$~=~$\vert\theta\vert$/$T_{1}$) \cite{Ramirez2001} are tiny in both compounds, although we cannot completely rule out a possibility of a sizable magnetic entropy below 2~K. Note that highly frustrated magnetic materials usually reveal a large frustration factor, $f$~\cite{Balents2010}.

\begin{figure} [t]
\includegraphics[width=\linewidth]{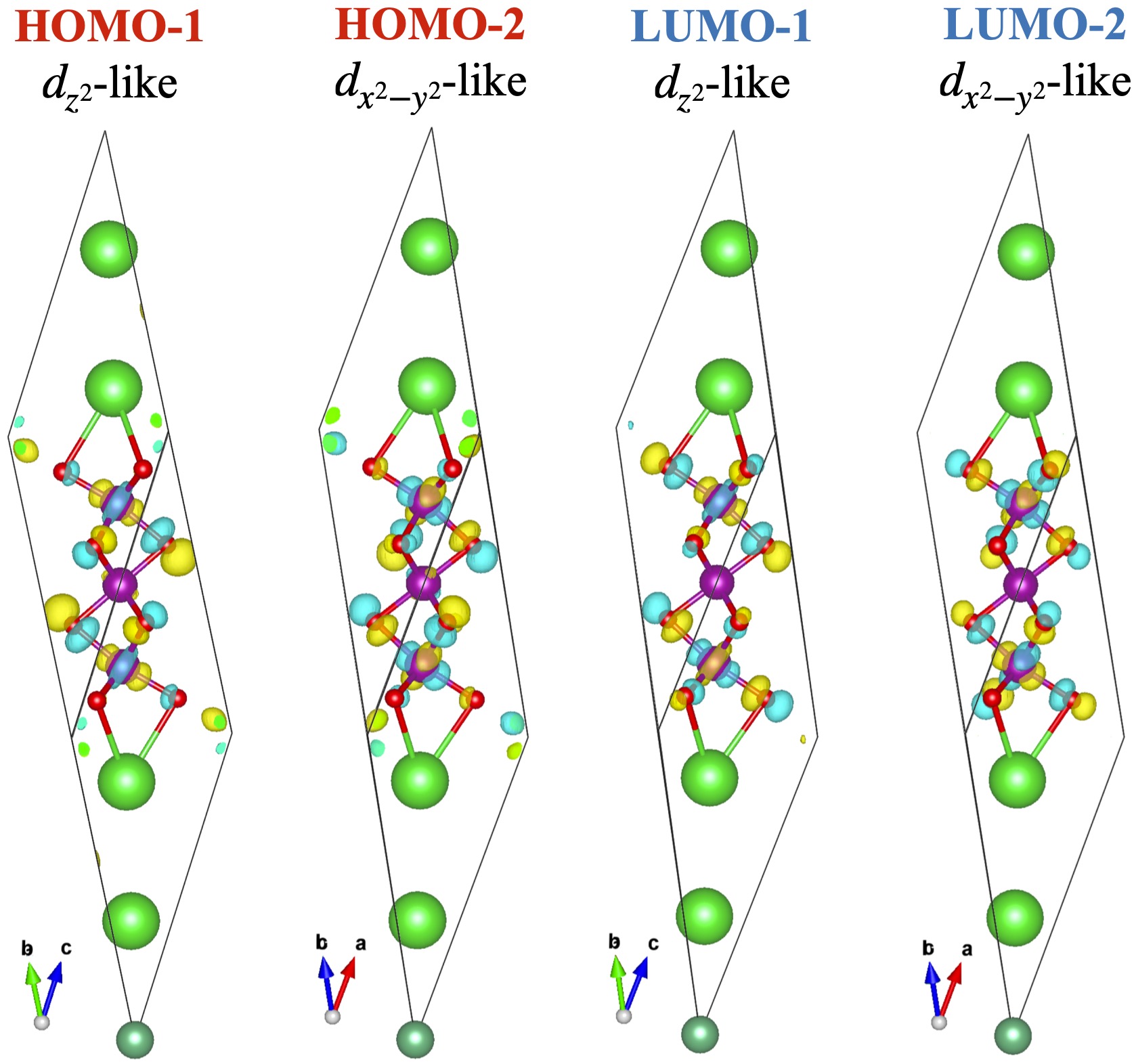}
\caption{HOMO (highest energy occupied molecular orbital) and LUMO (lowest energy unoccupied molecular orbital) wave functions of \BTMO at the $\Gamma$ point in Fig.~\ref{DFT_dos_band}(b). Violet and red spheres are Mn and and O ions, respectively. Ta ions are not shown as there is practically no contribution in the wave functions.
}
\label{DFT_wavefunctions}	
\end{figure}

As a more feasible explanation, we point out that the magnetic entropy of the itinerant magnet~\cite{Wohlfarth1977} is much smaller than that of the localized magnets. A tiny fraction of the magnetic entropy can be released via the magnetic transition in the itinerant magnet; for instance, only 2\% of the $S=1/2$ magnetic entropy ($\Delta$S$_{\rm mag}~=~0.02R$\rm{ln}2) was reported for the weakly ferromagnetic metal ZrZn$_{2}$~\cite{Wohlfarth1977, Viswanathan1975}. Hence it will be sensible to interpret that our heat capacity measurements detect the magnetic entropy dominantly released by the long-range magnetic order from the localized magnetic moment in both compounds. We note that Ref.~\cite{Nguyen2019} reported a puzzling magnetic entropy of \BNMO, which is much smaller than the values expected for both the $S=2$ and $S=3/2$ trimer. In our heat capacity measurements, we resolved this issue by using optimally grown polycrystalline samples.

Based on our results on \BXMO presented in this paper, further investigation of these peculiar magnetic properties would be highly desired. Pertaining to \BNMO, Ref.~\cite{Nguyen2019} suggested the presence of ordered magnetic moments inside the trimer even at 300~K; however, the long-range magnetic order stabilized by intertrimer exchange interactions appeared only at a lower temperature~\cite{Nguyen2019}. Thus neutron diffraction as a bulk probe will be useful to determine the magnetic structure and its evolution with temperature, which are related to the magnetic anomalies found in susceptibility measurements (Fig.~\ref{MTH}). On the other hand, muon spin relaxation as a local probe can measure the local magnetic field sensitively; for instance, it is suitable for studying static and dynamic magnetic properties of \BXMO. Thus the nature of the unusual paramagnetic state and the coexistence of localized and delocalized magnetism might be investigated by muons. These measurements could be sensitive to confirm the $S=3/2$ trimer in the localized moment picture.

Moreover, the underlying spin Hamiltonian for these compounds needs to be studied experimentally. In these compounds, spin dynamics can be determined by the inter- and intratrimer interaction of the Mn$_{3}$O$_{12}$ trimer because only Mn ions are magnetic (i.e., nonmagnetic Ba$^{2+}$ and Nb$^{5+}$ ions). A recent theory~\cite{Streltsov2018} suggested a ferromagnetic exchange in the $ab$ plane that is antiferromagnetically coupled along the $c$ axis. Consequently, inelastic neutron scattering (INS) will be crucial to assess the number of effectively dominant exchange couplings and their characteristics by testing various models, including the proposed one. As we found a strong coupling between Mn spins within the trimer from extracted effective moments at high temperatures [CW fits in Figs.~\ref{MTH}(c) and \ref{MTH}(d)], high-energy magnetic signals excited within the trimer are expected in INS measurements; for example, strong antiferromagnetic interactions between the middle and end Mn ions are anticipated in the AF2 model [Fig.~\ref{DFT_AF12}(b)]. Thus INS could be suitable to probe the $S=3/2$ trimer based on the energy and intensity of the trimer excitation.

Unlike the more studied 4$d$ and 5$d$ trimer-based materials~\cite{Nguyen2018, Nguyen2019_2, Thakur2020}, our results on \BXMO suggest an unconventional character---a partial molecular orbital state. Figure~\ref{DFT_wavefunctions} visualizes the wave functions of the partial molecular orbital state at $\Gamma$ point of \BTMO [Fig.~\ref{DFT_dos_band}(b)], demonstrating the absence of the $e_{g}$ electron in the central Mn ion. Also, it confirms that the electrons are not delocalized across the trimers via Ta/Nb $d$ orbitals. It is microscopically because the Ta/Nb orbitals are located at much higher energies, which makes the overlap ineffective [Fig.~\ref{DFT_dos_band}(b)].

The crucial difference for the different electronic structures of the 3$d$ and 4$d$/5$d$ compounds comes from the distinct hierarchy of the microscopic energy scales. In the 3$d$ case for \BXMOshort, the Coulomb interaction ($U$) is dominant. It is stronger than the intersite hybridization energy within the Mn trimer and the exchange splitting (Hund’s coupling) [Fig.~\ref{DFT_dos_band}(b)], which stabilizes the dominant local moment. This is a strongly localized scheme. The signs of $p$-orbital lobes surrounding the central Mn (in the overlap of ligand $p$ orbitals) are inversion antisymmetric for all four partial molecular orbital wave functions, effectively canceling inversion-symmetric $d$-orbital components at the central Mn ion. On the other hand, in the 4$d$/5$d$ counterparts \BNMeO, the intersite hybridization energy is dominant, stabilizing the molecular orbital state~\cite{Nguyen2018}.

This noteworthy difference in their natures of MO states implies a further research opportunity to examine the crossover between atomic and molecular descriptions of magnetism by mixing 3$d$ ions with heavier 4$d$ or 5$d$ ions; in particular, \BNRO~\cite{Nguyen2018} and \BNIO~\cite{Nguyen2019_2} can be synthesized. We note that the electronic structure and magnetism of \BXMOshort resemble those of the metallic double exchange in La$_{x}$Sr$_{1-x}$MnO$_{3}$~\cite{Park1998, Majumdar2014}, indicating that a possible ferromagnetic metal phase in \BXMOshort might be realized via an external magnetic field, pressure, or doping, which can be potentially useful in spintronics and magnetic devices.

Finally, in addition to \BXMO, a wide range of different compositions is possible in the similar hexagonal perovskite, Ba$_4$MM$'$$_3$O$_{12}$ (M~=~Nb, Ta, Ce, Pr and M$'$~=~Mn, Ru, Ir)~\cite{Fuentes2004, Nguyen2018, Shimoda2010, Shimoda2008, Shimoda2009} with face-sharing trimer octahedra. Therefore, this compound family is versatile but a less-studied platform for searching for unconventional magnetic properties, including molecular orbital-based magnetism~\cite{Wu2021, Nguyen2021}, calling for both experimental and theoretical studies in the future.

\section{Conclusions}
\label{sec:conclusions}
We reported an unconventional molecular orbital candidate \BTMO by comparing it to a recently reported compound \BNMO. We synthesized the polycrystalline sample by optimizing the amount of the MnO$_2$ starting powder monitored by x-ray diffraction measurements. Both magnetic susceptibility and heat capacity measurements revealed two compatible anomalies. Susceptibility showed the dominance of ferromagnetic intertrimer interaction. The effective magnetic moments indicated the strong coupling between spins within the trimer above the magnetic ordering temperature. The estimated magnetic entropy from heat capacity measurements is consistent with the antiferromagnetic $S = 3/2$ trimer with localized moments. These results indicate the combination of the localized $S = 3/2$ and one delocalized electron in the trimer. A\textit{b initio} calculations found that three localized electrons are present in each of the three Mn ions and one delocalized electron is spread over two-end Mn ions of the trimer, consistent with both susceptibility and heat capacity results. The magnetic ordering wave vector is predicted to be $q = (1/2, 0, 0)$. Thus our comprehensive results propose the partial molecular orbital in 3\textit{d} trimer-based materials—an unconventional electronic state with delocalized and localized electrons in a single compound. This could be understood by the competition between the hybrid interatomic orbitals within the Mn trimer and the local moment formation by on-site Coulomb correlations. In searching for novel electronic states, the possibility of finding the insulator-metal transition in this compound family by an external magnetic field, pressure, or doping warrants active further investigations in the future.

\section*{Acknowledgments}
This work was supported by the Institute for Basic Science (Grant No. IBS-R011-Y3) and Advanced Facility Center for Quantum Technology at Sungkyunkwan University. Part of this study has been performed using facilities at IBS Center for Correlated Electron Systems, Seoul National University. H.-S.K. acknowledges the support of the National Research Foundation of Korea (Basic Science Research Program, Grant No. 2020R1C1C1005900, RS-2023-00220471) and the support of computational resources, including technical assistance from the National Supercomputing Center of Korea (Grant No. KSC-2022-CRE-0358).

\bibliography{Manuscript}

\begin{thebibliography}{49}%
\makeatletter
\providecommand \@ifxundefined [1]{%
 \@ifx{#1\undefined}
}%
\providecommand \@ifnum [1]{%
 \ifnum #1\expandafter \@firstoftwo
 \else \expandafter \@secondoftwo
 \fi
}%
\providecommand \@ifx [1]{%
 \ifx #1\expandafter \@firstoftwo
 \else \expandafter \@secondoftwo
 \fi
}%
\providecommand \natexlab [1]{#1}%
\providecommand \enquote  [1]{``#1''}%
\providecommand \bibnamefont  [1]{#1}%
\providecommand \bibfnamefont [1]{#1}%
\providecommand \citenamefont [1]{#1}%
\providecommand \href@noop [0]{\@secondoftwo}%
\providecommand \href [0]{\begingroup \@sanitize@url \@href}%
\providecommand \@href[1]{\@@startlink{#1}\@@href}%
\providecommand \@@href[1]{\endgroup#1\@@endlink}%
\providecommand \@sanitize@url [0]{\catcode `\\12\catcode `\$12\catcode
  `\&12\catcode `\#12\catcode `\^12\catcode `\_12\catcode `\%12\relax}%
\providecommand \@@startlink[1]{}%
\providecommand \@@endlink[0]{}%
\providecommand \url  [0]{\begingroup\@sanitize@url \@url }%
\providecommand \@url [1]{\endgroup\@href {#1}{\urlprefix }}%
\providecommand \urlprefix  [0]{URL }%
\providecommand \Eprint [0]{\href }%
\providecommand \doibase [0]{https://doi.org/}%
\providecommand \selectlanguage [0]{\@gobble}%
\providecommand \bibinfo  [0]{\@secondoftwo}%
\providecommand \bibfield  [0]{\@secondoftwo}%
\providecommand \translation [1]{[#1]}%
\providecommand \BibitemOpen [0]{}%
\providecommand \bibitemStop [0]{}%
\providecommand \bibitemNoStop [0]{.\EOS\space}%
\providecommand \EOS [0]{\spacefactor3000\relax}%
\providecommand \BibitemShut  [1]{\csname bibitem#1\endcsname}%
\let\auto@bib@innerbib\@empty
\bibitem [{\citenamefont {Imada}\ \emph {et~al.}(1998)\citenamefont {Imada},
  \citenamefont {Fujimori},\ and\ \citenamefont {Tokura}}]{Imada1998}%
  \BibitemOpen
  \bibfield  {author} {\bibinfo {author} {\bibfnamefont {M.}~\bibnamefont
  {Imada}}, \bibinfo {author} {\bibfnamefont {A.}~\bibnamefont {Fujimori}},\
  and\ \bibinfo {author} {\bibfnamefont {Y.}~\bibnamefont {Tokura}},\
  }\bibfield  {title} {\bibinfo {title} {{Metal-insulator transitions}},\
  }\href {https://doi.org/10.1103/RevModPhys.70.1039} {\bibfield  {journal}
  {\bibinfo  {journal} {Rev. Mod. Phys.}\ }\textbf {\bibinfo {volume} {70}},\
  \bibinfo {pages} {1039} (\bibinfo {year} {1998})}\BibitemShut {NoStop}%
\bibitem [{\citenamefont {Streltsov}\ and\ \citenamefont
  {Khomskii}(2017)}]{Streltsov2017}%
  \BibitemOpen
  \bibfield  {author} {\bibinfo {author} {\bibfnamefont {S.~V.}\ \bibnamefont
  {Streltsov}}\ and\ \bibinfo {author} {\bibfnamefont {D.~I.}\ \bibnamefont
  {Khomskii}},\ }\bibfield  {title} {\bibinfo {title} {{Orbital physics in
  transition metal compounds: new trends}},\ }\href
  {https://doi.org/10.3367/ufne.2017.08.038196} {\bibfield  {journal} {\bibinfo
   {journal} {Phys.-Usp.}\ }\textbf {\bibinfo {volume} {60}},\ \bibinfo {pages}
  {1121} (\bibinfo {year} {2017})}\BibitemShut {NoStop}%
\bibitem [{\citenamefont {Klein}\ \emph {et~al.}(2011)\citenamefont {Klein},
  \citenamefont {Rousse}, \citenamefont {Damay}, \citenamefont {Porcher},
  \citenamefont {Andr\'e},\ and\ \citenamefont {Terasaki}}]{Klein2011}%
  \BibitemOpen
  \bibfield  {author} {\bibinfo {author} {\bibfnamefont {Y.}~\bibnamefont
  {Klein}}, \bibinfo {author} {\bibfnamefont {G.}~\bibnamefont {Rousse}},
  \bibinfo {author} {\bibfnamefont {F.}~\bibnamefont {Damay}}, \bibinfo
  {author} {\bibfnamefont {F.}~\bibnamefont {Porcher}}, \bibinfo {author}
  {\bibfnamefont {G.}~\bibnamefont {Andr\'e}},\ and\ \bibinfo {author}
  {\bibfnamefont {I.}~\bibnamefont {Terasaki}},\ }\bibfield  {title} {\bibinfo
  {title} {{Antiferromagnetic order and consequences on the transport
  properties of {B}a$_{4}${R}u$_{3}${O}$_{10}$}},\ }\href
  {https://doi.org/10.1103/PhysRevB.84.054439} {\bibfield  {journal} {\bibinfo
  {journal} {Phys. Rev. B}\ }\textbf {\bibinfo {volume} {84}},\ \bibinfo
  {pages} {054439} (\bibinfo {year} {2011})}\BibitemShut {NoStop}%
\bibitem [{\citenamefont {Streltsov}\ and\ \citenamefont
  {Khomskii}(2012)}]{Streltsov2012}%
  \BibitemOpen
  \bibfield  {author} {\bibinfo {author} {\bibfnamefont {S.~V.}\ \bibnamefont
  {Streltsov}}\ and\ \bibinfo {author} {\bibfnamefont {D.~I.}\ \bibnamefont
  {Khomskii}},\ }\bibfield  {title} {\bibinfo {title} {{Unconventional
  magnetism as a consequence of the charge disproportionation and the molecular
  orbital formation in {B}a$_{4}${R}u$_{3}${O}$_{10}$}},\ }\href
  {https://doi.org/10.1103/PhysRevB.86.064429} {\bibfield  {journal} {\bibinfo
  {journal} {Phys. Rev. B}\ }\textbf {\bibinfo {volume} {86}},\ \bibinfo
  {pages} {064429} (\bibinfo {year} {2012})}\BibitemShut {NoStop}%
\bibitem [{\citenamefont {Ye}\ \emph {et~al.}(2018)\citenamefont {Ye},
  \citenamefont {Kim}, \citenamefont {Kim}, \citenamefont {Won}, \citenamefont
  {Haule}, \citenamefont {Vanderbilt}, \citenamefont {Cheong},\ and\
  \citenamefont {Blumberg}}]{Ye2018}%
  \BibitemOpen
  \bibfield  {author} {\bibinfo {author} {\bibfnamefont {M.}~\bibnamefont
  {Ye}}, \bibinfo {author} {\bibfnamefont {H.-S.}\ \bibnamefont {Kim}},
  \bibinfo {author} {\bibfnamefont {J.-W.}\ \bibnamefont {Kim}}, \bibinfo
  {author} {\bibfnamefont {C.-J.}\ \bibnamefont {Won}}, \bibinfo {author}
  {\bibfnamefont {K.}~\bibnamefont {Haule}}, \bibinfo {author} {\bibfnamefont
  {D.}~\bibnamefont {Vanderbilt}}, \bibinfo {author} {\bibfnamefont {S.-W.}\
  \bibnamefont {Cheong}},\ and\ \bibinfo {author} {\bibfnamefont
  {G.}~\bibnamefont {Blumberg}},\ }\bibfield  {title} {\bibinfo {title}
  {{Covalency-driven collapse of strong spin-orbit coupling in face-sharing
  iridium octahedra}},\ }\href {https://doi.org/10.1103/PhysRevB.98.201105}
  {\bibfield  {journal} {\bibinfo  {journal} {Phys. Rev. B}\ }\textbf {\bibinfo
  {volume} {98}},\ \bibinfo {pages} {201105} (\bibinfo {year}
  {2018})}\BibitemShut {NoStop}%
\bibitem [{\citenamefont {Georges}\ \emph {et~al.}(2013)\citenamefont
  {Georges}, \citenamefont {Medici},\ and\ \citenamefont
  {Mravlje}}]{Georges2013}%
  \BibitemOpen
  \bibfield  {author} {\bibinfo {author} {\bibfnamefont {A.}~\bibnamefont
  {Georges}}, \bibinfo {author} {\bibfnamefont {L.~d.}\ \bibnamefont
  {Medici}},\ and\ \bibinfo {author} {\bibfnamefont {J.}~\bibnamefont
  {Mravlje}},\ }\bibfield  {title} {\bibinfo {title} {{Strong Correlations from
  Hund’s Coupling}},\ }\href
  {https://doi.org/10.1146/annurev-conmatphys-020911-125045} {\bibfield
  {journal} {\bibinfo  {journal} {Annu. Rev. Condens}\ }\textbf {\bibinfo
  {volume} {4}},\ \bibinfo {pages} {137} (\bibinfo {year} {2013})}\BibitemShut
  {NoStop}%
\bibitem [{\citenamefont {Nguyen}\ \emph {et~al.}(2019)\citenamefont {Nguyen},
  \citenamefont {Kong},\ and\ \citenamefont {Cava}}]{Nguyen2019}%
  \BibitemOpen
  \bibfield  {author} {\bibinfo {author} {\bibfnamefont {L.~T.}\ \bibnamefont
  {Nguyen}}, \bibinfo {author} {\bibfnamefont {T.}~\bibnamefont {Kong}},\ and\
  \bibinfo {author} {\bibfnamefont {R.~J.}\ \bibnamefont {Cava}},\ }\bibfield
  {title} {\bibinfo {title} {{Trimers of {Mn}{O}$_{6}$ octahedra and
  ferrimagnetism of {Ba$_{4}$NbMn$_{3}$O$_{12}$}}},\ }\href
  {https://doi.org/10.1088/2053-1591/ab0695} {\bibfield  {journal} {\bibinfo
  {journal} {Mater. Res. Express}\ }\textbf {\bibinfo {volume} {6}},\ \bibinfo
  {pages} {056108} (\bibinfo {year} {2019})}\BibitemShut {NoStop}%
\bibitem [{\citenamefont {Poeppelmeier}\ \emph {et~al.}(1982)\citenamefont
  {Poeppelmeier}, \citenamefont {Leonowicz}, \citenamefont {Scanlon},
  \citenamefont {Longo},\ and\ \citenamefont {Yelon}}]{Poeppelmeier1982}%
  \BibitemOpen
  \bibfield  {author} {\bibinfo {author} {\bibfnamefont {K.}~\bibnamefont
  {Poeppelmeier}}, \bibinfo {author} {\bibfnamefont {M.}~\bibnamefont
  {Leonowicz}}, \bibinfo {author} {\bibfnamefont {J.}~\bibnamefont {Scanlon}},
  \bibinfo {author} {\bibfnamefont {J.}~\bibnamefont {Longo}},\ and\ \bibinfo
  {author} {\bibfnamefont {W.}~\bibnamefont {Yelon}},\ }\bibfield  {title}
  {\bibinfo {title} {{Structure determination of {CaMnO$_3$} and
  {CaMnO$_{2.5}$} by {X}-ray and neutron methods}},\ }\href
  {https://doi.org/https://doi.org/10.1016/0022-4596(82)90292-4} {\bibfield
  {journal} {\bibinfo  {journal} {J. Solid State Chem.}\ }\textbf {\bibinfo
  {volume} {45}},\ \bibinfo {pages} {71} (\bibinfo {year} {1982})}\BibitemShut
  {NoStop}%
\bibitem [{MnM()}]{MnMetal}%
  \BibitemOpen
  \href@noop {} {\bibinfo {title} {{The Materials Project}}},\ \bibinfo
  {howpublished}
  {\url{https://materialsproject.org/materials/mp-8634/}}\BibitemShut {NoStop}%
\bibitem [{\citenamefont {Petříček}\ \emph {et~al.}(2014)\citenamefont
  {Petříček}, \citenamefont {Dušek},\ and\ \citenamefont
  {Palatinus}}]{Petricek2014}%
  \BibitemOpen
  \bibfield  {author} {\bibinfo {author} {\bibfnamefont {V.}~\bibnamefont
  {Petříček}}, \bibinfo {author} {\bibfnamefont {M.}~\bibnamefont
  {Dušek}},\ and\ \bibinfo {author} {\bibfnamefont {L.}~\bibnamefont
  {Palatinus}},\ }\bibfield  {title} {\bibinfo {title} {{Crystallographic
  Computing System JANA2006: General features}},\ }\href
  {https://doi.org/doi:10.1515/zkri-2014-1737} {\bibfield  {journal} {\bibinfo
  {journal} {Z. Kristallogr. Cryst. Mater.}\ }\textbf {\bibinfo {volume}
  {229}},\ \bibinfo {pages} {345} (\bibinfo {year} {2014})}\BibitemShut
  {NoStop}%
\bibitem [{\citenamefont {Kresse}\ and\ \citenamefont
  {Hafner}(1993)}]{Kresse1993}%
  \BibitemOpen
  \bibfield  {author} {\bibinfo {author} {\bibfnamefont {G.}~\bibnamefont
  {Kresse}}\ and\ \bibinfo {author} {\bibfnamefont {J.}~\bibnamefont
  {Hafner}},\ }\bibfield  {title} {\bibinfo {title} {{\textit{Ab initio}
  molecular dynamics for liquid metals}},\ }\href
  {https://doi.org/10.1103/PhysRevB.47.558} {\bibfield  {journal} {\bibinfo
  {journal} {Phys. Rev. B}\ }\textbf {\bibinfo {volume} {47}},\ \bibinfo
  {pages} {558} (\bibinfo {year} {1993})}\BibitemShut {NoStop}%
\bibitem [{\citenamefont {Kresse}\ and\ \citenamefont
  {Furthm\"uller}(1996)}]{Kresse1996}%
  \BibitemOpen
  \bibfield  {author} {\bibinfo {author} {\bibfnamefont {G.}~\bibnamefont
  {Kresse}}\ and\ \bibinfo {author} {\bibfnamefont {J.}~\bibnamefont
  {Furthm\"uller}},\ }\bibfield  {title} {\bibinfo {title} {{Efficient
  iterative schemes for \textit{ab initio} total-energy calculations using a
  plane-wave basis set}},\ }\href {https://doi.org/10.1103/PhysRevB.54.11169}
  {\bibfield  {journal} {\bibinfo  {journal} {Phys. Rev. B}\ }\textbf {\bibinfo
  {volume} {54}},\ \bibinfo {pages} {11169} (\bibinfo {year}
  {1996})}\BibitemShut {NoStop}%
\bibitem [{\citenamefont {Dudarev}\ \emph {et~al.}(1998)\citenamefont
  {Dudarev}, \citenamefont {Botton}, \citenamefont {Savrasov}, \citenamefont
  {Humphreys},\ and\ \citenamefont {Sutton}}]{Dudarev1998}%
  \BibitemOpen
  \bibfield  {author} {\bibinfo {author} {\bibfnamefont {S.~L.}\ \bibnamefont
  {Dudarev}}, \bibinfo {author} {\bibfnamefont {G.~A.}\ \bibnamefont {Botton}},
  \bibinfo {author} {\bibfnamefont {S.~Y.}\ \bibnamefont {Savrasov}}, \bibinfo
  {author} {\bibfnamefont {C.~J.}\ \bibnamefont {Humphreys}},\ and\ \bibinfo
  {author} {\bibfnamefont {A.~P.}\ \bibnamefont {Sutton}},\ }\bibfield  {title}
  {\bibinfo {title} {{Electron-energy-loss spectra and the structural stability
  of nickel oxide: An {LSDA+U} study}},\ }\href
  {https://doi.org/10.1103/PhysRevB.57.1505} {\bibfield  {journal} {\bibinfo
  {journal} {Phys. Rev. B}\ }\textbf {\bibinfo {volume} {57}},\ \bibinfo
  {pages} {1505} (\bibinfo {year} {1998})}\BibitemShut {NoStop}%
\bibitem [{\citenamefont {Lee}\ and\ \citenamefont {Han}(2013)}]{Alex2013}%
  \BibitemOpen
  \bibfield  {author} {\bibinfo {author} {\bibfnamefont {A.~T.}\ \bibnamefont
  {Lee}}\ and\ \bibinfo {author} {\bibfnamefont {M.~J.}\ \bibnamefont {Han}},\
  }\bibfield  {title} {\bibinfo {title} {{Charge transfer, confinement, and
  ferromagnetism in {LaMnO$_{3}$/LaNiO$_{3}$} (001) superlattices}},\ }\href
  {https://doi.org/10.1103/PhysRevB.88.035126} {\bibfield  {journal} {\bibinfo
  {journal} {Phys. Rev. B}\ }\textbf {\bibinfo {volume} {88}},\ \bibinfo
  {pages} {035126} (\bibinfo {year} {2013})}\BibitemShut {NoStop}%
\bibitem [{\citenamefont {Streltsov}\ and\ \citenamefont
  {Khomskii}(2018)}]{Streltsov2018}%
  \BibitemOpen
  \bibfield  {author} {\bibinfo {author} {\bibfnamefont {S.~V.}\ \bibnamefont
  {Streltsov}}\ and\ \bibinfo {author} {\bibfnamefont {D.~I.}\ \bibnamefont
  {Khomskii}},\ }\bibfield  {title} {\bibinfo {title} {{Cluster Magnetism of
  {Ba$_{4}$NbMn$_{3}$O$_{12}$}: Localized Electrons or Molecular Orbitals?}},\
  }\href {https://doi.org/10.1134/S0021364018220071} {\bibfield  {journal}
  {\bibinfo  {journal} {Jetp Lett.}\ }\textbf {\bibinfo {volume} {108}},\
  \bibinfo {pages} {686} (\bibinfo {year} {2018})}\BibitemShut {NoStop}%
\bibitem [{\citenamefont {Kim}\ \emph {et~al.}(2019)\citenamefont {Kim},
  \citenamefont {Haule},\ and\ \citenamefont {Vanderbilt}}]{HSK2019}%
  \BibitemOpen
  \bibfield  {author} {\bibinfo {author} {\bibfnamefont {H.-S.}\ \bibnamefont
  {Kim}}, \bibinfo {author} {\bibfnamefont {K.}~\bibnamefont {Haule}},\ and\
  \bibinfo {author} {\bibfnamefont {D.}~\bibnamefont {Vanderbilt}},\ }\bibfield
   {title} {\bibinfo {title} {{Mott Metal-Insulator Transitions in Pressurized
  Layered Trichalcogenides}},\ }\href
  {https://doi.org/10.1103/PhysRevLett.123.236401} {\bibfield  {journal}
  {\bibinfo  {journal} {Phys. Rev. Lett.}\ }\textbf {\bibinfo {volume} {123}},\
  \bibinfo {pages} {236401} (\bibinfo {year} {2019})}\BibitemShut {NoStop}%
\bibitem [{\citenamefont {Harms}\ \emph {et~al.}(2020)\citenamefont {Harms},
  \citenamefont {Kim}, \citenamefont {Clune}, \citenamefont {Smith},
  \citenamefont {O’Neal}, \citenamefont {Haglund}, \citenamefont {Mandrus},
  \citenamefont {Liu}, \citenamefont {Haule}, \citenamefont {Vanderbilt},\ and\
  \citenamefont {Musfeldt}}]{Harms2020}%
  \BibitemOpen
  \bibfield  {author} {\bibinfo {author} {\bibfnamefont {N.~C.}\ \bibnamefont
  {Harms}}, \bibinfo {author} {\bibfnamefont {H.-S.}\ \bibnamefont {Kim}},
  \bibinfo {author} {\bibfnamefont {A.~J.}\ \bibnamefont {Clune}}, \bibinfo
  {author} {\bibfnamefont {K.~A.}\ \bibnamefont {Smith}}, \bibinfo {author}
  {\bibfnamefont {K.~R.}\ \bibnamefont {O’Neal}}, \bibinfo {author}
  {\bibfnamefont {A.~V.}\ \bibnamefont {Haglund}}, \bibinfo {author}
  {\bibfnamefont {D.~G.}\ \bibnamefont {Mandrus}}, \bibinfo {author}
  {\bibfnamefont {Z.}~\bibnamefont {Liu}}, \bibinfo {author} {\bibfnamefont
  {K.}~\bibnamefont {Haule}}, \bibinfo {author} {\bibfnamefont
  {D.}~\bibnamefont {Vanderbilt}},\ and\ \bibinfo {author} {\bibfnamefont
  {J.~L.}\ \bibnamefont {Musfeldt}},\ }\bibfield  {title} {\bibinfo {title}
  {{Piezochromism in the magnetic chalcogenide MnPS$_{3}$}},\ }\href
  {https://doi.org/10.1038/s41535-020-00259-5} {\bibfield  {journal} {\bibinfo
  {journal} {npj Quantum Mater.}\ }\textbf {\bibinfo {volume} {5}},\ \bibinfo
  {pages} {56} (\bibinfo {year} {2020})}\BibitemShut {NoStop}%
\bibitem [{\citenamefont {Ali}\ \emph {et~al.}(2016)\citenamefont {Ali},
  \citenamefont {Kremer},\ and\ \citenamefont {Johnsson}}]{Ali2016}%
  \BibitemOpen
  \bibfield  {author} {\bibinfo {author} {\bibfnamefont {S.~I.}\ \bibnamefont
  {Ali}}, \bibinfo {author} {\bibfnamefont {R.~K.}\ \bibnamefont {Kremer}},\
  and\ \bibinfo {author} {\bibfnamefont {M.}~\bibnamefont {Johnsson}},\
  }\bibfield  {title} {\bibinfo {title} {{Hydrothermal Synthesis and Magnetic
  Characterization of the Quaternary Oxide
  {C}o{M}o$_{2}${S}b$_{2}${O}$_{10}$}},\ }\href
  {https://doi.org/10.1021/acs.inorgchem.6b02031} {\bibfield  {journal}
  {\bibinfo  {journal} {Inorg. Chem.}\ }\textbf {\bibinfo {volume} {55}},\
  \bibinfo {pages} {11490} (\bibinfo {year} {2016})}\BibitemShut {NoStop}%
\bibitem [{\citenamefont {Mugiraneza}\ and\ \citenamefont
  {Hallas}(2022)}]{Mugiraneza2022}%
  \BibitemOpen
  \bibfield  {author} {\bibinfo {author} {\bibfnamefont {S.}~\bibnamefont
  {Mugiraneza}}\ and\ \bibinfo {author} {\bibfnamefont {A.~M.}\ \bibnamefont
  {Hallas}},\ }\bibfield  {title} {\bibinfo {title} {{Tutorial: a beginner’s
  guide to interpreting magnetic susceptibility data with the Curie-Weiss
  law}},\ }\href {https://www.nature.com/articles/s42005-022-00853-y}
  {\bibfield  {journal} {\bibinfo  {journal} {Communications Physics}\ }\textbf
  {\bibinfo {volume} {5}},\ \bibinfo {pages} {95} (\bibinfo {year}
  {2022})}\BibitemShut {NoStop}%
\bibitem [{\citenamefont {Carlin}(1986)}]{Carlin1986}%
  \BibitemOpen
  \bibfield  {author} {\bibinfo {author} {\bibfnamefont {R.~L.}\ \bibnamefont
  {Carlin}},\ }\href {https://doi.org/10.1007/978-3-642-70733-9_2} {\emph
  {\bibinfo {title} {Magnetochemistry}}}\ (\bibinfo  {publisher} {Springer
  Berlin Heidelberg},\ \bibinfo {year} {1986})\BibitemShut {NoStop}%
\bibitem [{\citenamefont {Nguyen}\ \emph {et~al.}(2018)\citenamefont {Nguyen},
  \citenamefont {Halloran}, \citenamefont {Xie}, \citenamefont {Kong},
  \citenamefont {Broholm},\ and\ \citenamefont {Cava}}]{Nguyen2018}%
  \BibitemOpen
  \bibfield  {author} {\bibinfo {author} {\bibfnamefont {L.~T.}\ \bibnamefont
  {Nguyen}}, \bibinfo {author} {\bibfnamefont {T.}~\bibnamefont {Halloran}},
  \bibinfo {author} {\bibfnamefont {W.}~\bibnamefont {Xie}}, \bibinfo {author}
  {\bibfnamefont {T.}~\bibnamefont {Kong}}, \bibinfo {author} {\bibfnamefont
  {C.~L.}\ \bibnamefont {Broholm}},\ and\ \bibinfo {author} {\bibfnamefont
  {R.~J.}\ \bibnamefont {Cava}},\ }\bibfield  {title} {\bibinfo {title}
  {{Geometrically frustrated trimer-based Mott insulator}},\ }\href
  {https://doi.org/10.1103/PhysRevMaterials.2.054414} {\bibfield  {journal}
  {\bibinfo  {journal} {Phys. Rev. Mater.}\ }\textbf {\bibinfo {volume} {2}},\
  \bibinfo {pages} {054414} (\bibinfo {year} {2018})}\BibitemShut {NoStop}%
\bibitem [{\citenamefont {Thakur}\ \emph {et~al.}(2020)\citenamefont {Thakur},
  \citenamefont {Chattopadhyay}, \citenamefont {Doert}, \citenamefont
  {Herrmannsdörfer},\ and\ \citenamefont {Felser}}]{Thakur2020}%
  \BibitemOpen
  \bibfield  {author} {\bibinfo {author} {\bibfnamefont {G.~S.}\ \bibnamefont
  {Thakur}}, \bibinfo {author} {\bibfnamefont {S.}~\bibnamefont
  {Chattopadhyay}}, \bibinfo {author} {\bibfnamefont {T.}~\bibnamefont
  {Doert}}, \bibinfo {author} {\bibfnamefont {T.}~\bibnamefont
  {Herrmannsdörfer}},\ and\ \bibinfo {author} {\bibfnamefont {C.}~\bibnamefont
  {Felser}},\ }\bibfield  {title} {\bibinfo {title} {{Crystal Growth of
  Spin-frustrated {Ba$_4$Nb$_{0.8}$Ir$_{3.2}$O$_{12}$}: A Possible Spin Liquid
  Material}},\ }\href {https://doi.org/10.1021/acs.cgd.0c00262} {\bibfield
  {journal} {\bibinfo  {journal} {Cryst. Growth Des.}\ }\textbf {\bibinfo
  {volume} {20}},\ \bibinfo {pages} {2871} (\bibinfo {year}
  {2020})}\BibitemShut {NoStop}%
\bibitem [{\citenamefont {Kittel}(2004)}]{Kittel2004}%
  \BibitemOpen
  \bibfield  {author} {\bibinfo {author} {\bibfnamefont {C.}~\bibnamefont
  {Kittel}},\ }\href
  {http://www.amazon.com/Introduction-Solid-Physics-Charles-Kittel/dp/047141526X/ref=dp_ob_title_bk}
  {\emph {\bibinfo {title} {{Introduction to Solid State Physics}}}}\ (\bibinfo
   {publisher} {John Wiley \& Sons, Ltd},\ \bibinfo {year} {2004})\BibitemShut
  {NoStop}%
\bibitem [{\citenamefont {Caslin}\ \emph {et~al.}(2014)\citenamefont {Caslin},
  \citenamefont {Kremer}, \citenamefont {Razavi}, \citenamefont {Schulz},
  \citenamefont {Mu\~noz}, \citenamefont {Pertlik}, \citenamefont {Liu},
  \citenamefont {Whangbo},\ and\ \citenamefont {Law}}]{Caslin2014}%
  \BibitemOpen
  \bibfield  {author} {\bibinfo {author} {\bibfnamefont {K.}~\bibnamefont
  {Caslin}}, \bibinfo {author} {\bibfnamefont {R.~K.}\ \bibnamefont {Kremer}},
  \bibinfo {author} {\bibfnamefont {F.~S.}\ \bibnamefont {Razavi}}, \bibinfo
  {author} {\bibfnamefont {A.}~\bibnamefont {Schulz}}, \bibinfo {author}
  {\bibfnamefont {A.}~\bibnamefont {Mu\~noz}}, \bibinfo {author} {\bibfnamefont
  {F.}~\bibnamefont {Pertlik}}, \bibinfo {author} {\bibfnamefont
  {J.}~\bibnamefont {Liu}}, \bibinfo {author} {\bibfnamefont {M.-H.}\
  \bibnamefont {Whangbo}},\ and\ \bibinfo {author} {\bibfnamefont {J.~M.}\
  \bibnamefont {Law}},\ }\bibfield  {title} {\bibinfo {title} {Characterization
  of the spin-$\frac{1}{2}$ linear-chain ferromagnet {{CuAs$_{2}$O$_{4}$}}},\
  }\href {https://doi.org/10.1103/PhysRevB.89.014412} {\bibfield  {journal}
  {\bibinfo  {journal} {Phys. Rev. B}\ }\textbf {\bibinfo {volume} {89}},\
  \bibinfo {pages} {014412} (\bibinfo {year} {2014})}\BibitemShut {NoStop}%
\bibitem [{\citenamefont {Sebastian}\ \emph {et~al.}(2021)\citenamefont
  {Sebastian}, \citenamefont {Somesh}, \citenamefont {Nandi}, \citenamefont
  {Ahmed}, \citenamefont {Bag}, \citenamefont {Baenitz}, \citenamefont {Koo},
  \citenamefont {Sichelschmidt}, \citenamefont {Tsirlin}, \citenamefont
  {Furukawa},\ and\ \citenamefont {Nath}}]{Sebastian2021}%
  \BibitemOpen
  \bibfield  {author} {\bibinfo {author} {\bibfnamefont {S.~J.}\ \bibnamefont
  {Sebastian}}, \bibinfo {author} {\bibfnamefont {K.}~\bibnamefont {Somesh}},
  \bibinfo {author} {\bibfnamefont {M.}~\bibnamefont {Nandi}}, \bibinfo
  {author} {\bibfnamefont {N.}~\bibnamefont {Ahmed}}, \bibinfo {author}
  {\bibfnamefont {P.}~\bibnamefont {Bag}}, \bibinfo {author} {\bibfnamefont
  {M.}~\bibnamefont {Baenitz}}, \bibinfo {author} {\bibfnamefont
  {B.}~\bibnamefont {Koo}}, \bibinfo {author} {\bibfnamefont {J.}~\bibnamefont
  {Sichelschmidt}}, \bibinfo {author} {\bibfnamefont {A.~A.}\ \bibnamefont
  {Tsirlin}}, \bibinfo {author} {\bibfnamefont {Y.}~\bibnamefont {Furukawa}},\
  and\ \bibinfo {author} {\bibfnamefont {R.}~\bibnamefont {Nath}},\ }\bibfield
  {title} {\bibinfo {title} {{Quasi-one-dimensional magnetism in the
  spin-$\frac{1}{2}$ antiferromagnet
  {${\mathrm{BaNa}}_{2}\mathrm{Cu}{({\mathrm{VO}}_{4})}_{2}$}}},\ }\href
  {https://doi.org/10.1103/PhysRevB.103.064413} {\bibfield  {journal} {\bibinfo
   {journal} {Phys. Rev. B}\ }\textbf {\bibinfo {volume} {103}},\ \bibinfo
  {pages} {064413} (\bibinfo {year} {2021})}\BibitemShut {NoStop}%
\bibitem [{\citenamefont {Setyawan}\ and\ \citenamefont
  {Curtarolo}(2010)}]{Setyawan2010}%
  \BibitemOpen
  \bibfield  {author} {\bibinfo {author} {\bibfnamefont {W.}~\bibnamefont
  {Setyawan}}\ and\ \bibinfo {author} {\bibfnamefont {S.}~\bibnamefont
  {Curtarolo}},\ }\bibfield  {title} {\bibinfo {title} {{High-throughput
  electronic band structure calculations: Challenges and tools}},\ }\href
  {https://doi.org/https://doi.org/10.1016/j.commatsci.2010.05.010} {\bibfield
  {journal} {\bibinfo  {journal} {Comput. Mater. Sci.}\ }\textbf {\bibinfo
  {volume} {49}},\ \bibinfo {pages} {299} (\bibinfo {year} {2010})}\BibitemShut
  {NoStop}%
\bibitem [{\citenamefont {White}(1969)}]{White1969}%
  \BibitemOpen
  \bibfield  {author} {\bibinfo {author} {\bibfnamefont {R.~L.}\ \bibnamefont
  {White}},\ }\bibfield  {title} {\bibinfo {title} {{R}eview of {R}ecent {W}ork
  on the {M}agnetic and {S}pectroscopic {P}roperties of the {R}are-{E}arth
  {O}rthoferrites},\ }\href {https://doi.org/10.1063/1.1657530} {\bibfield
  {journal} {\bibinfo  {journal} {Journal of Applied Physics}\ }\textbf
  {\bibinfo {volume} {40}},\ \bibinfo {pages} {1061} (\bibinfo {year}
  {1969})}\BibitemShut {NoStop}%
\bibitem [{\citenamefont {Young}\ \emph {et~al.}(2012)\citenamefont {Young},
  \citenamefont {Chapon},\ and\ \citenamefont {Petrenko}}]{Young2012}%
  \BibitemOpen
  \bibfield  {author} {\bibinfo {author} {\bibfnamefont {O.}~\bibnamefont
  {Young}}, \bibinfo {author} {\bibfnamefont {L.~C.}\ \bibnamefont {Chapon}},\
  and\ \bibinfo {author} {\bibfnamefont {O.~A.}\ \bibnamefont {Petrenko}},\
  }\bibfield  {title} {\bibinfo {title} {{Low temperature magnetic structure of
  geometrically frustrated {S}r{H}o$_{2}${O}$_{4}$}},\ }\href
  {https://dx.doi.org/10.1088/1742-6596/391/1/012081} {\bibfield  {journal}
  {\bibinfo  {journal} {Journal of Physics: Conference Series}\ }\textbf
  {\bibinfo {volume} {391}},\ \bibinfo {pages} {012081} (\bibinfo {year}
  {2012})}\BibitemShut {NoStop}%
\bibitem [{\citenamefont {Casto}\ \emph {et~al.}(2015)\citenamefont {Casto},
  \citenamefont {Clune}, \citenamefont {Yokosuk}, \citenamefont {Musfeldt},
  \citenamefont {Williams}, \citenamefont {Zhuang}, \citenamefont {Lin},
  \citenamefont {Xiao}, \citenamefont {Hennig}, \citenamefont {Sales},
  \citenamefont {Yan},\ and\ \citenamefont {Mandrus}}]{Casto2015}%
  \BibitemOpen
  \bibfield  {author} {\bibinfo {author} {\bibfnamefont {L.~D.}\ \bibnamefont
  {Casto}}, \bibinfo {author} {\bibfnamefont {A.~J.}\ \bibnamefont {Clune}},
  \bibinfo {author} {\bibfnamefont {M.~O.}\ \bibnamefont {Yokosuk}}, \bibinfo
  {author} {\bibfnamefont {J.~L.}\ \bibnamefont {Musfeldt}}, \bibinfo {author}
  {\bibfnamefont {T.~J.}\ \bibnamefont {Williams}}, \bibinfo {author}
  {\bibfnamefont {H.~L.}\ \bibnamefont {Zhuang}}, \bibinfo {author}
  {\bibfnamefont {M.-W.}\ \bibnamefont {Lin}}, \bibinfo {author} {\bibfnamefont
  {K.}~\bibnamefont {Xiao}}, \bibinfo {author} {\bibfnamefont {R.~G.}\
  \bibnamefont {Hennig}}, \bibinfo {author} {\bibfnamefont {B.~C.}\
  \bibnamefont {Sales}}, \bibinfo {author} {\bibfnamefont {J.-Q.}\ \bibnamefont
  {Yan}},\ and\ \bibinfo {author} {\bibfnamefont {D.}~\bibnamefont {Mandrus}},\
  }\bibfield  {title} {\bibinfo {title} {{Strong spin-lattice coupling in
  {C}r{S}i{T}e$_{3}$}},\ }\href {https://doi.org/10.1063/1.4914134} {\bibfield
  {journal} {\bibinfo  {journal} {APL Mater.}\ }\textbf {\bibinfo {volume}
  {3}},\ \bibinfo {pages} {041515} (\bibinfo {year} {2015})}\BibitemShut
  {NoStop}%
\bibitem [{\citenamefont {Moriya}\ and\ \citenamefont
  {Takahashi}(1984)}]{Moriya1984}%
  \BibitemOpen
  \bibfield  {author} {\bibinfo {author} {\bibfnamefont {T.}~\bibnamefont
  {Moriya}}\ and\ \bibinfo {author} {\bibfnamefont {Y.}~\bibnamefont
  {Takahashi}},\ }\bibfield  {title} {\bibinfo {title} {{Itinerant Electron
  Magnetism}},\ }\href
  {https://www.annualreviews.org/doi/abs/10.1146/annurev.ms.14.080184.000245}
  {\bibfield  {journal} {\bibinfo  {journal} {Annual Review of Materials
  Science}\ }\textbf {\bibinfo {volume} {14}},\ \bibinfo {pages} {1} (\bibinfo
  {year} {1984})}\BibitemShut {NoStop}%
\bibitem [{\citenamefont {Moriya}(1985)}]{Moriya1985}%
  \BibitemOpen
  \bibfield  {author} {\bibinfo {author} {\bibfnamefont {T.}~\bibnamefont
  {Moriya}},\ }\href {https://doi.org/10.1007/978-3-642-82499-9} {\emph
  {\bibinfo {title} {{Spin Fluctuations in Itinerant Electron Magnetism}}}}\
  (\bibinfo  {publisher} {Springer Berlin Heidelberg},\ \bibinfo {year}
  {1985})\BibitemShut {NoStop}%
\bibitem [{\citenamefont {du~T.~de Lacheisserie}\ \emph
  {et~al.}(2002)\citenamefont {du~T.~de Lacheisserie}, \citenamefont
  {Gignoux},\ and\ \citenamefont {Schlenker}}]{Lacheisserie2002}%
  \BibitemOpen
  \bibfield  {author} {\bibinfo {author} {\bibfnamefont {{\'E}.}~\bibnamefont
  {du~T.~de Lacheisserie}}, \bibinfo {author} {\bibfnamefont {D.}~\bibnamefont
  {Gignoux}},\ and\ \bibinfo {author} {\bibfnamefont {M.}~\bibnamefont
  {Schlenker}},\ }\href {https://doi.org/10.1007/978-0-387-23062-7_8} {\emph
  {\bibinfo {title} {{Magnetism: Fundamentals, Materials and Applications}}}}\
  (\bibinfo  {publisher} {Springer New York},\ \bibinfo {year}
  {2002})\BibitemShut {NoStop}%
\bibitem [{\citenamefont {Moriya}(1991)}]{Moriya1991}%
  \BibitemOpen
  \bibfield  {author} {\bibinfo {author} {\bibfnamefont {T.}~\bibnamefont
  {Moriya}},\ }\bibfield  {title} {\bibinfo {title} {{Theory of itinerant
  electron magnetism}},\ }\href
  {https://doi.org/https://doi.org/10.1016/0304-8853(91)90824-T} {\bibfield
  {journal} {\bibinfo  {journal} {Journal of Magnetism and Magnetic Materials}\
  }\textbf {\bibinfo {volume} {100}},\ \bibinfo {pages} {261} (\bibinfo {year}
  {1991})}\BibitemShut {NoStop}%
\bibitem [{\citenamefont {Topping}\ and\ \citenamefont
  {Blundell}(2019)}]{Topping2019}%
  \BibitemOpen
  \bibfield  {author} {\bibinfo {author} {\bibfnamefont {C.~V.}\ \bibnamefont
  {Topping}}\ and\ \bibinfo {author} {\bibfnamefont {S.~J.}\ \bibnamefont
  {Blundell}},\ }\bibfield  {title} {\bibinfo {title} {{A.C. susceptibility as
  a probe of low-frequency magnetic dynamics}},\ }\href
  {https://doi.org/10.1088/1361-648X/aaed96} {\bibfield  {journal} {\bibinfo
  {journal} {Journal of Physics: Condensed Matter}\ }\textbf {\bibinfo {volume}
  {31}},\ \bibinfo {pages} {013001} (\bibinfo {year} {2019})}\BibitemShut
  {NoStop}%
\bibitem [{\citenamefont {Zhong}\ \emph {et~al.}(2019)\citenamefont {Zhong},
  \citenamefont {Guo}, \citenamefont {Xu}, \citenamefont {Xu},\ and\
  \citenamefont {Cava}}]{Zhong2019}%
  \BibitemOpen
  \bibfield  {author} {\bibinfo {author} {\bibfnamefont {R.}~\bibnamefont
  {Zhong}}, \bibinfo {author} {\bibfnamefont {S.}~\bibnamefont {Guo}}, \bibinfo
  {author} {\bibfnamefont {G.}~\bibnamefont {Xu}}, \bibinfo {author}
  {\bibfnamefont {Z.}~\bibnamefont {Xu}},\ and\ \bibinfo {author}
  {\bibfnamefont {R.~J.}\ \bibnamefont {Cava}},\ }\bibfield  {title} {\bibinfo
  {title} {{Strong quantum fluctuations in a quantum spin liquid candidate with
  a {C}o-based triangular lattice}},\ }\href
  {https://doi.org/10.1073/pnas.1906483116} {\bibfield  {journal} {\bibinfo
  {journal} {Proc. Natl. Acad. Sci. U.S.A.}\ }\textbf {\bibinfo {volume}
  {116}},\ \bibinfo {pages} {14505} (\bibinfo {year} {2019})}\BibitemShut
  {NoStop}%
\bibitem [{\citenamefont {Ramirez}(2001)}]{Ramirez2001}%
  \BibitemOpen
  \bibfield  {author} {\bibinfo {author} {\bibfnamefont {A.}~\bibnamefont
  {Ramirez}},\ }\href@noop {} {\emph {\bibinfo {title} {{Handbook of Magnetic
  Materials}}}},\ Vol.~\bibinfo {volume} {13}\ (\bibinfo  {publisher} {Gulf
  Professional Publishing, Houston, TX},\ \bibinfo {year} {2001})\
  Chap.~\bibinfo {chapter} {4}, p.\ \bibinfo {pages} {423}\BibitemShut
  {NoStop}%
\bibitem [{\citenamefont {Balents}(2010)}]{Balents2010}%
  \BibitemOpen
  \bibfield  {author} {\bibinfo {author} {\bibfnamefont {L.}~\bibnamefont
  {Balents}},\ }\bibfield  {title} {\bibinfo {title} {{S}pin liquids in
  frustrated magnets},\ }\href
  {https://doi.org/https://doi.org/10.1038/nature08917} {\bibfield  {journal}
  {\bibinfo  {journal} {Nature}\ }\textbf {\bibinfo {volume} {464}},\ \bibinfo
  {pages} {199} (\bibinfo {year} {2010})}\BibitemShut {NoStop}%
\bibitem [{\citenamefont {Wohlfarth}(1977)}]{Wohlfarth1977}%
  \BibitemOpen
  \bibfield  {author} {\bibinfo {author} {\bibfnamefont {E.~P.}\ \bibnamefont
  {Wohlfarth}},\ }\bibfield  {title} {\bibinfo {title} {{Thermodynamic aspects
  of itinerant electron magnetism}},\ }\href
  {https://doi.org/https://doi.org/10.1016/0378-4363(77)90199-1} {\bibfield
  {journal} {\bibinfo  {journal} {Physica B+C}\ }\textbf {\bibinfo {volume}
  {91}},\ \bibinfo {pages} {305} (\bibinfo {year} {1977})}\BibitemShut
  {NoStop}%
\bibitem [{\citenamefont {Viswanathan}\ and\ \citenamefont
  {Clinton}(1975)}]{Viswanathan1975}%
  \BibitemOpen
  \bibfield  {author} {\bibinfo {author} {\bibfnamefont {R.}~\bibnamefont
  {Viswanathan}}\ and\ \bibinfo {author} {\bibfnamefont {J.~R.}\ \bibnamefont
  {Clinton}},\ }\bibfield  {title} {\bibinfo {title} {{Magnetic entropy and
  magnetization of {Z}r{Z}n$_{2}$}},\ }\href
  {https://doi.org/https://doi.org/10.1063/1.29924} {\bibfield  {journal}
  {\bibinfo  {journal} {AIP Conference Proceedings}\ }\textbf {\bibinfo
  {volume} {24}},\ \bibinfo {pages} {416} (\bibinfo {year} {1975})}\BibitemShut
  {NoStop}%
\bibitem [{\citenamefont {Nguyen}\ and\ \citenamefont
  {Cava}(2019)}]{Nguyen2019_2}%
  \BibitemOpen
  \bibfield  {author} {\bibinfo {author} {\bibfnamefont {L.~T.}\ \bibnamefont
  {Nguyen}}\ and\ \bibinfo {author} {\bibfnamefont {R.~J.}\ \bibnamefont
  {Cava}},\ }\bibfield  {title} {\bibinfo {title} {{Trimer-based spin liquid
  candidate
  {$\mathrm{B}{\mathrm{a}}_{4}\mathrm{NbI}{\mathrm{r}}_{3}{\mathrm{O}}_{12}$}}},\
  }\href {https://doi.org/10.1103/PhysRevMaterials.3.014412} {\bibfield
  {journal} {\bibinfo  {journal} {Phys. Rev. Mater.}\ }\textbf {\bibinfo
  {volume} {3}},\ \bibinfo {pages} {014412} (\bibinfo {year}
  {2019})}\BibitemShut {NoStop}%
\bibitem [{\citenamefont {Park}\ \emph {et~al.}(1998)\citenamefont {Park},
  \citenamefont {Vescovo}, \citenamefont {Kim}, \citenamefont {Kwon},
  \citenamefont {Ramesh},\ and\ \citenamefont {Venkatesan}}]{Park1998}%
  \BibitemOpen
  \bibfield  {author} {\bibinfo {author} {\bibfnamefont {J.-H.}\ \bibnamefont
  {Park}}, \bibinfo {author} {\bibfnamefont {E.}~\bibnamefont {Vescovo}},
  \bibinfo {author} {\bibfnamefont {H.-J.}\ \bibnamefont {Kim}}, \bibinfo
  {author} {\bibfnamefont {C.}~\bibnamefont {Kwon}}, \bibinfo {author}
  {\bibfnamefont {R.}~\bibnamefont {Ramesh}},\ and\ \bibinfo {author}
  {\bibfnamefont {T.}~\bibnamefont {Venkatesan}},\ }\bibfield  {title}
  {\bibinfo {title} {{Direct evidence for a half-metallic ferromagnet}},\
  }\href {https://doi.org/10.1038/33883} {\bibfield  {journal} {\bibinfo
  {journal} {Nature}\ }\textbf {\bibinfo {volume} {392}},\ \bibinfo {pages}
  {794} (\bibinfo {year} {1998})}\BibitemShut {NoStop}%
\bibitem [{\citenamefont {Majumdar}\ and\ \citenamefont {van
  Dijken}(2014)}]{Majumdar2014}%
  \BibitemOpen
  \bibfield  {author} {\bibinfo {author} {\bibfnamefont {S.}~\bibnamefont
  {Majumdar}}\ and\ \bibinfo {author} {\bibfnamefont {S.}~\bibnamefont {van
  Dijken}},\ }\bibfield  {title} {\bibinfo {title} {{Pulsed laser deposition of
  {La$_{1-x}$Sr$_{x}$MnO$_{3}$}: thin-film properties and spintronic
  applications}},\ }\href {https://doi.org/10.1088/0022-3727/47/3/034010}
  {\bibfield  {journal} {\bibinfo  {journal} {J. Phys. D: Appl. Phys.}\
  }\textbf {\bibinfo {volume} {47}},\ \bibinfo {pages} {034010} (\bibinfo
  {year} {2014})}\BibitemShut {NoStop}%
\bibitem [{\citenamefont {Fuentes}\ \emph {et~al.}(2004)\citenamefont
  {Fuentes}, \citenamefont {Boulahya},\ and\ \citenamefont
  {Amador}}]{Fuentes2004}%
  \BibitemOpen
  \bibfield  {author} {\bibinfo {author} {\bibfnamefont {A.~F.}\ \bibnamefont
  {Fuentes}}, \bibinfo {author} {\bibfnamefont {K.}~\bibnamefont {Boulahya}},\
  and\ \bibinfo {author} {\bibfnamefont {U.}~\bibnamefont {Amador}},\
  }\bibfield  {title} {\bibinfo {title} {{Novel rare-earth-containing
  manganites {Ba$_{4}$REMn$_{3}$O$_{12}$} {(RE=Ce, Pr) with 12R} structure}},\
  }\href {https://doi.org/https://doi.org/10.1016/j.jssc.2003.08.025}
  {\bibfield  {journal} {\bibinfo  {journal} {J. Solid State Chem.}\ }\textbf
  {\bibinfo {volume} {177}},\ \bibinfo {pages} {714} (\bibinfo {year}
  {2004})}\BibitemShut {NoStop}%
\bibitem [{\citenamefont {Shimoda}\ \emph {et~al.}(2010)\citenamefont
  {Shimoda}, \citenamefont {Doi}, \citenamefont {Wakeshima},\ and\
  \citenamefont {Hinatsu}}]{Shimoda2010}%
  \BibitemOpen
  \bibfield  {author} {\bibinfo {author} {\bibfnamefont {Y.}~\bibnamefont
  {Shimoda}}, \bibinfo {author} {\bibfnamefont {Y.}~\bibnamefont {Doi}},
  \bibinfo {author} {\bibfnamefont {M.}~\bibnamefont {Wakeshima}},\ and\
  \bibinfo {author} {\bibfnamefont {Y.}~\bibnamefont {Hinatsu}},\ }\bibfield
  {title} {\bibinfo {title} {{Magnetic and electrical properties of quadruple
  perovskites with 12 layer structures {Ba$_4$LnM$_3$O$_{12}$ (Ln=Rare earths;
  M=Ru, Ir)}: The role of metal–metal bonding in perovskite-related
  oxides}},\ }\href
  {https://doi.org/https://doi.org/10.1016/j.jssc.2010.06.023} {\bibfield
  {journal} {\bibinfo  {journal} {J. Solid State Chem.}\ }\textbf {\bibinfo
  {volume} {183}},\ \bibinfo {pages} {1962} (\bibinfo {year}
  {2010})}\BibitemShut {NoStop}%
\bibitem [{\citenamefont {Shimoda}\ \emph {et~al.}(2008)\citenamefont
  {Shimoda}, \citenamefont {Doi}, \citenamefont {Hinatsu},\ and\ \citenamefont
  {Ohoyama}}]{Shimoda2008}%
  \BibitemOpen
  \bibfield  {author} {\bibinfo {author} {\bibfnamefont {Y.}~\bibnamefont
  {Shimoda}}, \bibinfo {author} {\bibfnamefont {Y.}~\bibnamefont {Doi}},
  \bibinfo {author} {\bibfnamefont {Y.}~\bibnamefont {Hinatsu}},\ and\ \bibinfo
  {author} {\bibfnamefont {K.}~\bibnamefont {Ohoyama}},\ }\bibfield  {title}
  {\bibinfo {title} {{Synthesis, Crystal Structures, and Magnetic Properties of
  New 12{L}-Perovskites {Ba$_4$LnRu$_3$O$_{12}$ (Ln = Lanthanides)}}},\ }\href
  {https://doi.org/10.1021/cm800708g} {\bibfield  {journal} {\bibinfo
  {journal} {Chem. Mater.}\ }\textbf {\bibinfo {volume} {20}},\ \bibinfo
  {pages} {4512} (\bibinfo {year} {2008})}\BibitemShut {NoStop}%
\bibitem [{\citenamefont {Shimoda}\ \emph {et~al.}(2009)\citenamefont
  {Shimoda}, \citenamefont {Doi}, \citenamefont {Wakeshima},\ and\
  \citenamefont {Hinatsu}}]{Shimoda2009}%
  \BibitemOpen
  \bibfield  {author} {\bibinfo {author} {\bibfnamefont {Y.}~\bibnamefont
  {Shimoda}}, \bibinfo {author} {\bibfnamefont {Y.}~\bibnamefont {Doi}},
  \bibinfo {author} {\bibfnamefont {M.}~\bibnamefont {Wakeshima}},\ and\
  \bibinfo {author} {\bibfnamefont {Y.}~\bibnamefont {Hinatsu}},\ }\bibfield
  {title} {\bibinfo {title} {{Synthesis and magnetic properties of
  {12L-perovskites Ba$_4$LnIr$_3$O$_{12}$ (Ln=Lanthanides)}}},\ }\href
  {https://doi.org/https://doi.org/10.1016/j.jssc.2009.07.056} {\bibfield
  {journal} {\bibinfo  {journal} {J. Solid State Chem.}\ }\textbf {\bibinfo
  {volume} {182}},\ \bibinfo {pages} {2873} (\bibinfo {year}
  {2009})}\BibitemShut {NoStop}%
\bibitem [{\citenamefont {Wu}\ \emph {et~al.}(2021)\citenamefont {Wu},
  \citenamefont {Yan}, \citenamefont {Guo}, \citenamefont {Wang}, \citenamefont
  {Yin},\ and\ \citenamefont {Kuang}}]{Wu2021}%
  \BibitemOpen
  \bibfield  {author} {\bibinfo {author} {\bibfnamefont {J.}~\bibnamefont
  {Wu}}, \bibinfo {author} {\bibfnamefont {X.}~\bibnamefont {Yan}}, \bibinfo
  {author} {\bibfnamefont {W.}~\bibnamefont {Guo}}, \bibinfo {author}
  {\bibfnamefont {X.}~\bibnamefont {Wang}}, \bibinfo {author} {\bibfnamefont
  {C.}~\bibnamefont {Yin}},\ and\ \bibinfo {author} {\bibfnamefont
  {X.}~\bibnamefont {Kuang}},\ }\bibfield  {title} {\bibinfo {title}
  {{Molecule-like cluster magnetism and cationic order in the new hexagonal
  perovskite {Ba$_4$Sn$_{1.1}$Mn$_{2.9}$O$_{12}$}}},\ }\href
  {https://doi.org/10.1039/D1RA07841K} {\bibfield  {journal} {\bibinfo
  {journal} {RSC Adv.}\ }\textbf {\bibinfo {volume} {11}},\ \bibinfo {pages}
  {40235} (\bibinfo {year} {2021})}\BibitemShut {NoStop}%
\bibitem [{\citenamefont {Nguyen}\ and\ \citenamefont
  {Cava}(2021)}]{Nguyen2021}%
  \BibitemOpen
  \bibfield  {author} {\bibinfo {author} {\bibfnamefont {L.~T.}\ \bibnamefont
  {Nguyen}}\ and\ \bibinfo {author} {\bibfnamefont {R.~J.}\ \bibnamefont
  {Cava}},\ }\bibfield  {title} {\bibinfo {title} {{Hexagonal Perovskites as
  Quantum Materials}},\ }\href {https://doi.org/10.1021/acs.chemrev.0c00622}
  {\bibfield  {journal} {\bibinfo  {journal} {Chem. Rev.}\ }\textbf {\bibinfo
  {volume} {121}},\ \bibinfo {pages} {2935} (\bibinfo {year}
  {2021})}\BibitemShut {NoStop}%
\bibitem [{\citenamefont {Furness}\ \emph {et~al.}(2020)\citenamefont
  {Furness}, \citenamefont {Kaplan}, \citenamefont {Ning}, \citenamefont
  {Perdew},\ and\ \citenamefont {Sun}}]{Furness2020}%
  \BibitemOpen
  \bibfield  {author} {\bibinfo {author} {\bibfnamefont {J.~W.}\ \bibnamefont
  {Furness}}, \bibinfo {author} {\bibfnamefont {A.~D.}\ \bibnamefont {Kaplan}},
  \bibinfo {author} {\bibfnamefont {J.}~\bibnamefont {Ning}}, \bibinfo {author}
  {\bibfnamefont {J.~P.}\ \bibnamefont {Perdew}},\ and\ \bibinfo {author}
  {\bibfnamefont {J.}~\bibnamefont {Sun}},\ }\bibfield  {title} {\bibinfo
  {title} {{Accurate and Numerically Efficient r2SCAN Meta-Generalized Gradient
  Approximation}},\ }\href
  {https://doi.org/https://pubs.acs.org/doi/10.1021/acs.jpclett.0c02405}
  {\bibfield  {journal} {\bibinfo  {journal} {The Journal of Physical Chemistry
  Letters}\ }\textbf {\bibinfo {volume} {11}},\ \bibinfo {pages} {8208}
  (\bibinfo {year} {2020})}\BibitemShut {NoStop}%
\end{thebibliography}%

\appendix

\begin{figure} [t]
\includegraphics[width=\linewidth]{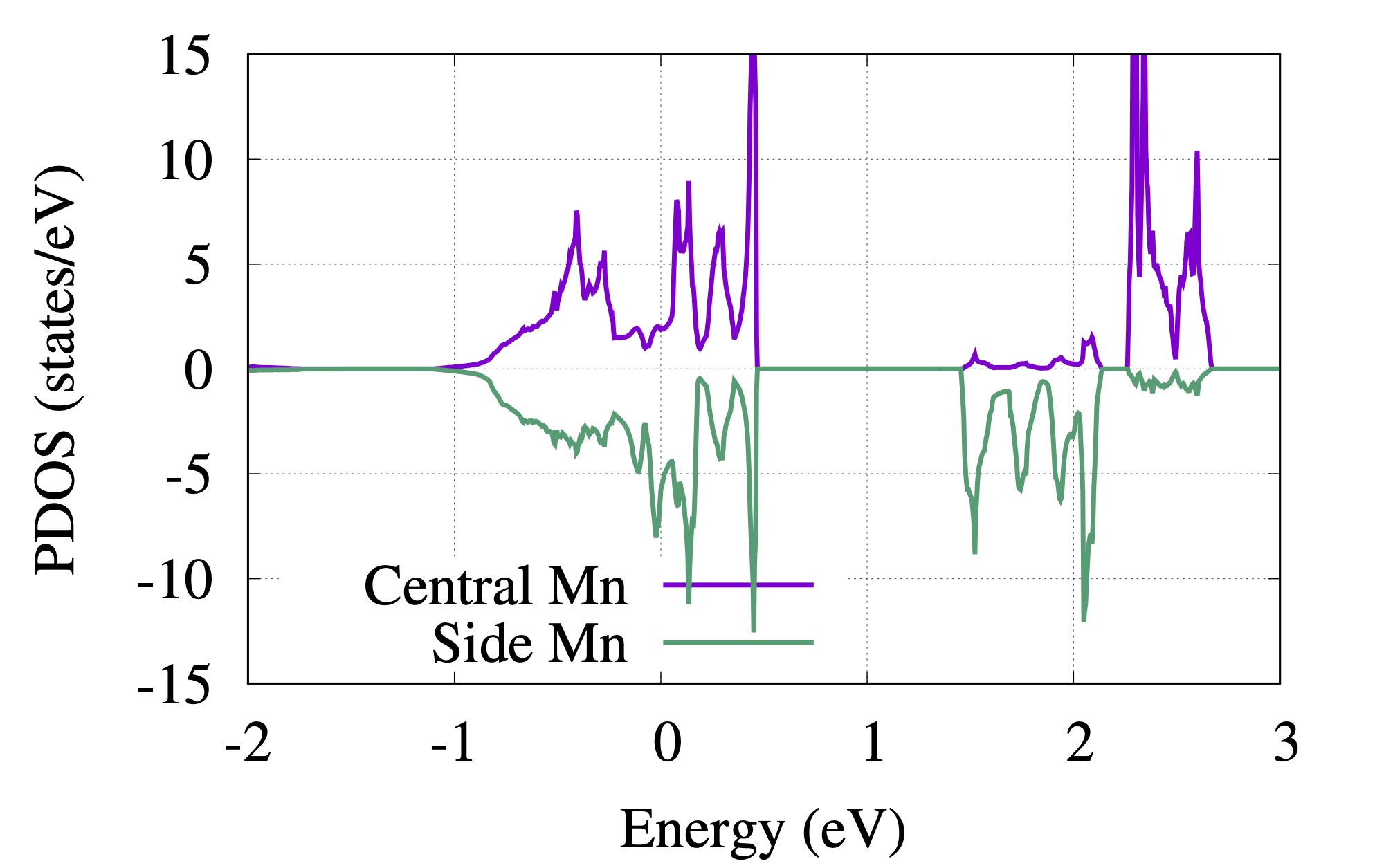}
\caption{Projected densities of states for central and side Mn atoms within the Mn trimer from nonmagnetic DFT calculations without $U_{\rm eff}$. Note that $t_{2g}$ orbital states are close to the Fermi level ($E$ = 0), and $e_{g}$ orbital states are located between 1.5 and 3 eV.
}
\label{DFT_nonmagnetic}	
\end{figure}

\section{Nonmagnetic DFT calculations}
\label{appendix:DFT}
To understand the physics above the magnetic transition temperature, we performed additional DFT calculations. Figure~\ref{DFT_nonmagnetic} shows Mn-site-projected densities of states from a nonmagnetic DFT calculation without $U_{\rm eff}$. Therein it can be seen that the $t_{2g}$-$e_{g}$ splitting in central and side Mn sites are about 2.5 and 1.8 eV, respectively, whereas the strength of intersite hybridization within the Mn trimer ($\sim$ the bandwidth of $t_{2g}$ and $e_{g}$ orbitals) is shown to be about 1 eV. We note that Fig.~\ref{DFT_dos_band}(b) presents that the size of the exchange splitting is about 4 eV when $U_{\rm eff}$ = 4 eV in our DFT+ $U_{\rm eff}$ calculations. A separate calculation employing a parameter-free r2SCAN meta-GGA functional~\cite{Furness2020} reveals the size of the exchange splitting to be about 4 eV, which is consistent with results in Fig.~\ref{DFT_dos_band}(b). Overall, the hierarchy of on-site energy scales at Mn sites is as follows; Coulomb interaction ($U$) $>$ exchange splitting (Hund’s coupling) $>$ cubic crystal fields ($t_{2g}$-$e_{g}$ splitting) $>$ intersite hybridization within Mn trimer. This relation stabilizes the partial molecular orbital state in \BXMO.

\end{document}